\documentclass[journal]{IEEEtran}
\usepackage{amsmath,amsfonts}
\usepackage{algorithmic}
\usepackage{algorithm}
\usepackage{array}
\usepackage[caption=false,font=normalsize,labelfont=sf,textfont=sf]{subfig}
\usepackage[colorlinks,linkcolor=red,anchorcolor=blue,citecolor=green]{hyperref}
\usepackage{textcomp}
\usepackage{stfloats}
\usepackage{url}
\usepackage{enumitem}
\usepackage{verbatim}
\usepackage{comment}
\usepackage{graphicx}
\usepackage{cite}
\usepackage{multirow}
\newcommand{\model}{MHNN}

\usepackage{xcolor}

\begin{document}

\title{Disentangling Imperfect: A  Wavelet-Infused Multilevel Heterogeneous Network for Human Activity Recognition in Flawed Wearable Sensor Data}

\author{Mengna Liu, Dong Xiang,  Xu Cheng, ~\IEEEmembership{Member,~IEEE,}  Xiufeng Liu, Dalin Zhang, Shengyong Chen, ~\IEEEmembership{Senior Member,~IEEE}, Christian S. Jensen, ~\IEEEmembership{IEEE Fellow}
 % stops a space

\thanks{Mengna Liu, Dong Xiang, Xu Cheng, Shengyong Chen are with the School of Computer Science and Engineering, Tianjin University of Technology, Tianjin, China.}% <-this % stops a space
\thanks{Xiufeng Liu is with the Department of Technology, Management, and Economics, Technical University of Denmark, Kongens Lyngby, Denmark.}
\thanks{Dalin Zhang and Christian S. Jensen are with the Department of Computer Science, Aalborg University, Aalborg, Denmark.}% <-this % stops a space
}

\markboth{IEEE Transactions on Mobile Computing}%
{Shell \MakeLowercase{\textit{et al.}}: A Sample Article Using IEEEtran.cls for IEEE Journals}


\maketitle

\begin{abstract}
The popularity and diffusion of wearable devices provides new opportunities for sensor-based human activity recognition that leverages deep learning-based algorithms. Although impressive advances have been made, two major challenges remain. First, sensor data is often incomplete or noisy due to sensor placement and other issues as well as data transmission failure, calling for imputation of missing values, which also introduces noise. Second, human activity has multi-scale characteristics. Thus, different groups of people and even the same person may behave differently under different circumstances. To address these challenges, we propose a multilevel heterogeneous neural network, called \model, for sensor data analysis. We utilize multilevel discrete wavelet decomposition to extract multi-resolution features from sensor data. This enables distinguishing signals with different frequencies, thereby suppressing noise. As the components resulting from the decomposition are heterogeneous, we equip the proposed model with heterogeneous feature extractors that enable the learning of multi-scale features. Due to the complementarity of these features, we also include a cross aggregation module for enhancing their interactions. An experimental study using seven publicly available datasets offers evidence that \model~can outperform other cutting-edge models and offers evidence of robustness to missing values and noise. An ablation study confirms the importance of each module.
\end{abstract}

\begin{IEEEkeywords}
Human activity recognition, deep learning, multilevel discrete wavelet transform, wearable sensors.
\end{IEEEkeywords}

\section{Introduction}
%\IEEEPARstart{H}{uman} activity recognition (HAR) is a significant research field within the Internet of Things (IoT), with widespread applications in many domains including smart homes \cite{Lina2018WITS}, intelligent traffic control \cite{2017Participatory} and intelligent surveillance systems \cite{2018A, qiu2022multi}. With the increasing prevalence of smart devices and wearable technologies, it has become feasible to monitor human behavior using signal collected from the built-in sensors.

\IEEEPARstart{H}{uman} activity recognition (HAR) is a dynamic research domain that strives to identify and categorize human activities through the utilization of sensors affixed to or integrated within the human body, such as accelerometers, gyroscopes, and magnetometers. HAR has many applications in various domains, such as smart healthcare \cite{yao2018wits}, sports \cite{2017Participatory}, entertainment \cite{2018A}, and security \cite{qiu2022multi}. HAR can be crafted to oversee individuals' physical activity and well-being status, to provide feedback and guidance for athletes, to create immersive and interactive gaming experiences, and to detect abnormal or suspicious behaviors.

However, HAR is a challenging task due to the complexity and diversity of human activities and the limitations and variability of sensor data. Human activities can be categorized into different levels of granularity, such as gestures, actions, activities, and events. Moreover, human activities can be performed by different people in various ways, with discrepant durations, frequencies, and intensities. Sensor data is susceptible to various factors, such as sensor placement, orientation, sampling rate, noise, drift, and occlusion. Therefore, designing robust and accurate HAR systems requires addressing several issues, such as sensor selection and fusion \cite{2018ActionRecog}, data preprocessing and segmentation \cite{2019StackedLSTM_HAR}, feature extraction and selection \cite{Ankita2017Human}, activity modeling and classification \cite{2017ResBiLSTM_HAR}, and evaluation and validation \cite{2018ANovelRandom}.

%Sensor-based HAR has gained considerable attention from researchers due to its convenience and no environmental constraints. Early approaches for action classification mainly focus on shallow machine learning (ML) algorithms, including support vector machine (SVM) \cite{Ankita2017Human}, random forest (RF) \cite{2018ANovelRandom} and Bayesian networks \cite{2018ActionRecog}. Sensor-specific and task-specific features are first extracted from the signal manually, and then are fed into ML model to identify activities. ML-based approaches make full use of human knowledge and have strong interpretability. However, the process of manual feature extraction is time-consuming and high cost. Furthermore, models based on manually extracted features can only achieve suboptimal performance due to the shallow nature of these features \cite{yang2015deep}.

In the past decade, sensor-based HAR has gained considerable attention, primarily owing to its convenience and absence of environmental constraints. Early approaches to activity classification mainly employ shallow machine learning (ML) algorithms, including support vector machines (SVM) \cite{Ankita2017Human}, random forests (RF) \cite{2018ANovelRandom}, and Bayesian networks \cite{2018ActionRecog}. 
%These methods require manual feature extraction from the sensor data based on domain knowledge and prior assumptions.
Sensor-specific and task-specific features are first extracted manually from signals, and are then fed into ML models to identify activity. ML-based approaches rely on domain knowledge and have strong interpretability. While, manual feature extraction is expensive and time-consuming. Furthermore, the performance of models based on manually extracted features is hampered by the shallow nature of such features \cite{yang2015deep}. 
During the past few years, deep learning (DL)-based models have have emerged as dominant in the field of sensor-based HAR and now exhibit remarkable performance. Since feature extraction is integrated into the DL model training process, extracted features are more abstract and task-relevant when compared to the manually extracted features. Preeminent among deep learning models for sensor-based HAR are convolutional neural networks (CNN), recurrent neural networks (RNN), and the Transformer architecture. CNNs excel at capturing spatial patterns and at hierarchically learning features from sub-sequences of the input \cite{yang2015deep,ha2016convolutional, tang2022multiscale}. RNNs are highly effective at modeling sequential dependencies \cite{2017ResBiLSTM_HAR, guan2017ensembles, 2019HAR_LSTM, 2019StackedLSTM_HAR}. The Transformer leverages an attention mechanism to enable critical time steps to play a stronger role in model prediction \cite{wang2019attention,mahmud2020HAR_CNN_attention}. Hybrid proposals combine multiple neural architectures to extract features with stronger representational power \cite{2021AMultibranch, xia2020lstm}. 

Although DL-based methods have achieved good performance, two challenges remain that need additional attention:
%Nevertheless, there are still some challenges remaining for DL-based methods:

\begin{figure}[!t]
\centering
\includegraphics[width=3in]{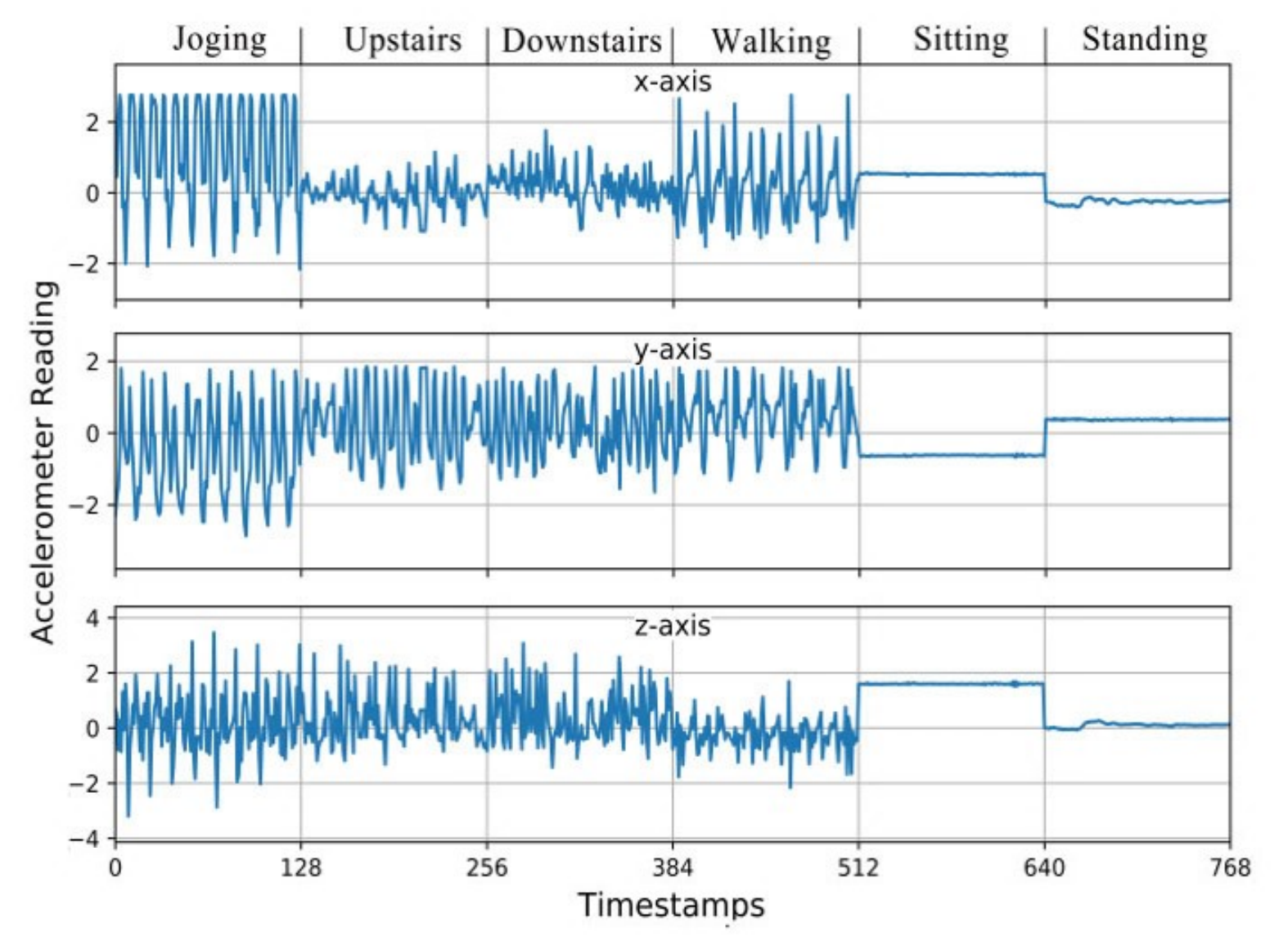}
\caption{Acceleration waveform spanning 2.56 seconds for each activity. \cite{xia2020lstm}}
\label{activity_signals}

\end{figure}

%Although the deep learning-based methods have achieved good performance, there are still some challenges remaining: 

%\begin{itemize}
 %   \item \textbf{Feature Extraction}: Fig. \ref{activity_signals} illustrates the acceleration waveform for each activity over a 2.56-second interval, totaling 128 data points, from the WISDM dataset. The visual representation reveals that different activities can exhibit varying time scales. For example, jogging signals show considerably higher amplitude and frequency fluctuations compared to walking signals. In contrast, sitting and standing activities produce relatively stable signals. Consequently, the challenge lies in how to devise a method to extract features at multiple resolutions. This would enable the model to recognize patterns across different time scales, and thereby differentiate between various activities.
  %  \item \textbf{Data Quality}: The quality of sensor data plays a crucial role in model development. Sensor data inevitably contain various forms of noise and missing values, which cannot be fully eradicated even with rigorous preprocessing techniques. Suboptimal data quality can adversely affect the performance of deep learning models, leading to a degradation in their predictive capabilities.
%\end{itemize}

\begin{itemize}[left=0pt]
\item \textbf{Incomplete and Noisy Data}: The quality of sensor data plays a crucial role in model development. Sensor data frequently contains various forms of noise and is incomplete, and this often cannot be addressed by even rigorous preprocessing. Suboptimal data quality can degrade the performance of DL-based models and lead to erroneous predictions. 

\item \textbf{Multi-scale Characteristic}: Fig. \ref{activity_signals} illustrates the acceleration waveforms corresponding to six activities over a 2.56-second interval, corresponding to 128 data points excerpted from the WISDM dataset. It reveals that different activities can exhibit varying time scales. For example, the jogging signals exhibit considerably higher amplitude and frequency fluctuations than do the walking signals. Further, sitting and standing activities produce relatively stable signals. Consequently, the challenge is how to devise a method to extract features at multiple resolutions. This would enable the model to recognize patterns across different time scales and thereby differentiate between different activities.

% \item \textbf{Data Quantity}: The quantity of sensor data is another important factor that affects model performance. DL models usually require large amounts of labeled data to train effectively and avoid overfitting. However, collecting and labeling sensor data is a labor-intensive and expensive process, especially for complex and diverse human activities. Therefore, developing methods that can achieve high accuracy with limited data is desirable. 
\end{itemize}

%To overcome the above challenges, we propose a multilevel heterogeneous neural network based on the multilevel discrete wavelet decomposition (MDWD) for HAR. MDWD can not only extract multi-resolution feature from time and frequency domains but also separate noise from signal due to the high frequency property of noise. With the deepening of the wavelet decomposition level, information quantity contained in the components decreases as the signal resolution decreases. If all the multi-resolution components use the same feature extractor, it will lead to insufficient extraction of high-resolution features and over-fitting of low-resolution features. So a heterogeneous feature learner is designed to extract representational features from multi-resolution components. The deeper components correspond to a simpler feature extraction subnet. As we all know, the high frequency component contains more details, and the low frequency component contains more approximate information. The high and low frequency components are complementary, so the interaction between the multilevel features is very useful for feature extraction, which is always neglected. We design a feature aggregation module to learn the interactive information between multilevel features. In short, the main contributions of this paper are as follows:

To address above challenges, we propose a novel method for HAR based on a Multilevel Heterogeneous Neural Network (\model) and multilevel discrete wavelet decomposition (MDWD). MDWD can decompose the original signal into multiple components with different resolutions in both time and frequency domains\cite{mallat1989theory}. Moreover, MDWD can separate noise from signal due to the high frequency property of noise. When deepening the wavelet decomposition level, information contained in the components decreases as the signal resolution decreases. %However, different components contain different amounts of information and require different types of feature extractors.
If all the multi-resolution components use the same feature extractor, this may lead to insufficient extraction of high-resolution features and over-fitting of low-resolution features. Thus, we design a heterogeneous feature extractor that adapts neural networks of varying complexity for different levels of components. The deeper components correspond to a simpler feature extraction subnet, such as a single convolutional layer or a fully connected layer. The shallower components correspond to a more complex feature extraction subnet, such as stacked CNN block or Transformer block. Moreover, we design a feature aggregation module that learns the integrated information between multilevel features. That is, the high frequency components contain more detail, where the low frequency components contain information that is more approximate. The high and low frequency components are complementary, making the interaction between the multilevel features useful for feature extraction.

In short, the main contributions of the paper are:
\begin{itemize}[left=0pt]
\item %We propose a multilevel heterogeneous deep learning framework, namely wavelet transform based multilevel heterogeneous convolutional neural network (\model), for sensor-based HAR. The original signal is decomposed by MDWD, which can not only extract multi-resolution signal components, but also de-noising. Heterogeneous feature extractor is designed to extract abstract features from each component respectively. The signal with deeper layers corresponds to a simpler feature extractor. A cross aggregation module is designed to extract cross feature between multi-resolution components.
The paper proposes a DL-based algorithm for sensor-based HAR based on \model~and MDWD, it can extract sub-signals at multiple resolutions and combine them in a hierarchical manner. The paper also introduces a cross aggregation module that can extract cross features between multilevel components.

\item %Extensive experiments are conducted to verify the effectiveness of \model. We compare our model with state-of-the-art methods on seven datasets and achieve supervisor performance. Missing ratio experiment and noise adding experiment demonstrate the robustness of our model. Ablation studies are provided to analysis the contribution of each module to the overall system. The optimum parameters of model were evaluated by sensitivity analysis.
This study assesses the effectiveness of the proposed approach across seven public HAR datasets. The findings suggest that our method outperforms 28 contemporary techniques relying on handcrafted features or deep learning model. The paper also demonstrates the robustness of the proposed method to noise and missing values. The influence of the parameters in the proposed model is verified by the sensitivity analysis. The paper also provides ablation studies to show the effectiveness of each component of the proposed method.

\end{itemize}
 
The rest of this paper is organized as follows. Section \ref{related_work} presents a brief description of recent signal-based HAR methods. Section \ref{method} describes the overall architecture of the proposed model \model. Section \ref{experiment} presents an experimental study. Finally, Section \ref{conclusion} concludes the paper.

\begin{figure*}[ht!]
\centering
\includegraphics[width=0.95\textwidth]{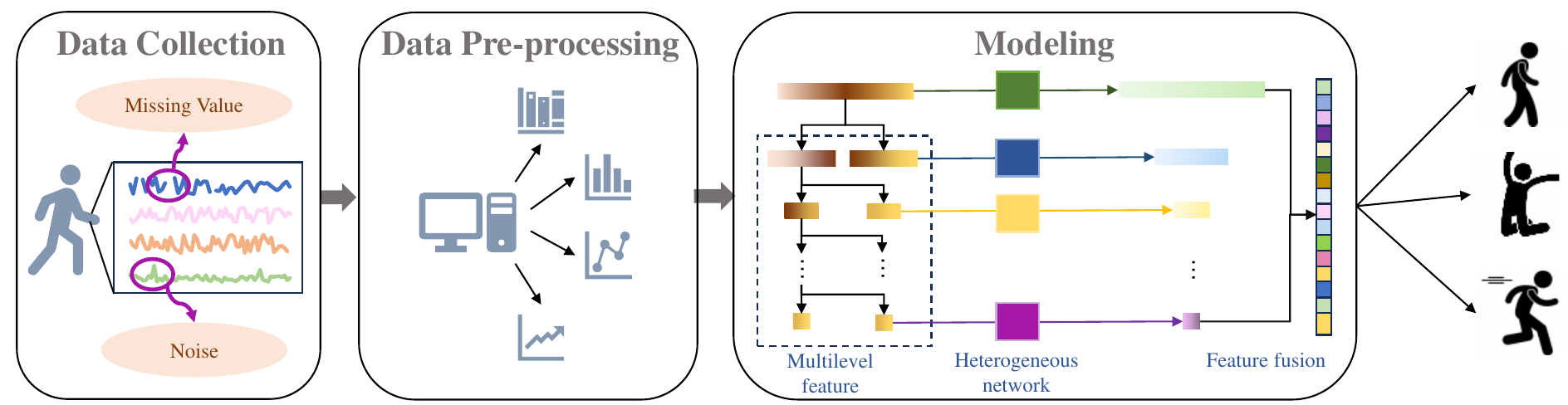}
\caption{The architecture of proposed model for sensor-based HAR.\label{whole_archi_ab}}
\end{figure*}

\begin{figure}[ht!]
\centering
\includegraphics[width=0.47\textwidth]{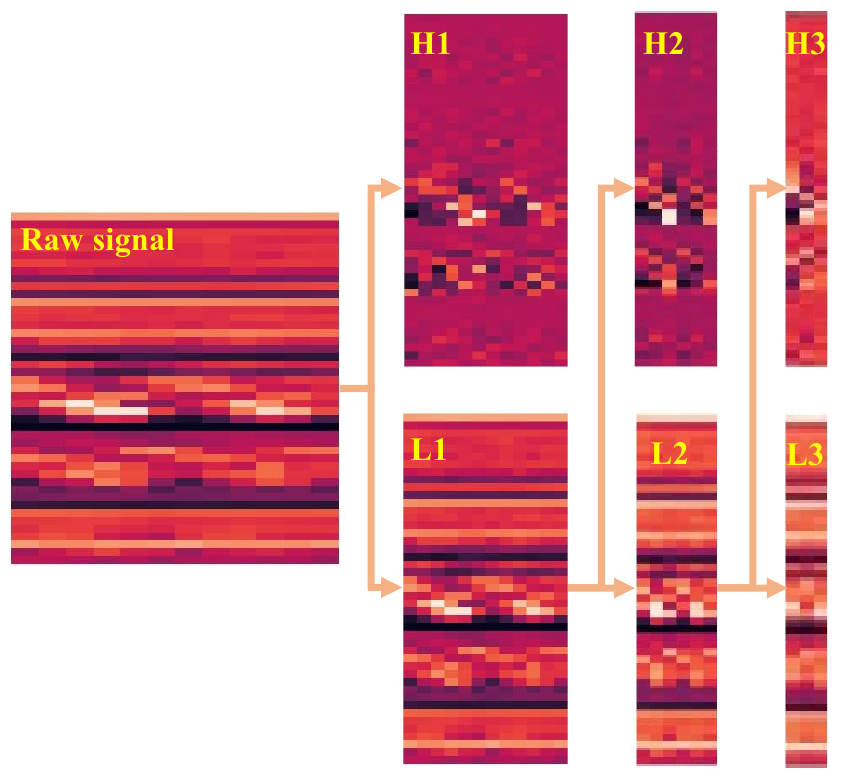}
\caption{Signal and the components decomposed by MDWD.}
\label{wavelet_visual}
\end{figure}

\section{RELATED WORK}
\label{related_work}
Sensor-based HAR has recently seen substantial progress. The traditional HAR approaches feed manually extracted features into shallow ML-based models to identify activities. Most manually extracted features consist of statistical information, such as median, variance, mean, and range. The ML models used for HAR include SVMs, RFs, and KNNs \cite{chen2017performance,chen2017robust}. An early study introduces SVMs for HAR task \cite{2012HumanSVM}. Nunes et al. \cite{2017A_rf} propose a differential evolution random forest algorithm to classify human actions. Tharwat et al. \cite{2018RecognizingKNN} employ KNNs to distinguish human activities. Particle swarm optimization is deployed to search for the optimal KNN parameter settings. The effectiveness of proposal is proved on ten datasets.

Although ML-based algorithms have achieved good performance on sensor-based HAR, the manual feature extraction approach requires feature engineers to have strong domain knowledge, making feature extraction expensive and time-consuming. DL-based methods can extract more abstract and powerful features, and can do so automatically. Yang et al. \cite{yang2015deep} introduce CNN to HAR. Their architecture use CNN to extract feature from time series data. Experiments on Opportunity dataset indicate better performance than previous sophisticated methods. Wang et al. \cite{wang2019attention} apply attention mechanism to a CNN to locate labeled activities in long sensor sequences. By selectively attending to important parts of a time series, the attention mechanisms can eliminate noise in other parts of the time series, leading to a more robust models. Yu et al. \cite{guan2017ensembles} propose an ensemble model of deep LSTM learners. Ensemble learning aims to be more robust and to achieve better accuracy than individual LSTM networks by training multiple sub-learners on different perspectives of the data. Zhao et al. \cite{2017ResBiLSTM_HAR} design the Deep-Res-Bidir-LSTM structure, it stacks four Bi-LSTM modules and adds a residual structure between the Bi-LSTM modules. Xia et al. \cite{xia2020lstm} propose a hybrid model combining LSTMs and CNNs to extract long-term dependencies information and local correlation simultaneously. AI-qaness et al. \cite{al2022multi} present a algorithm named Multi-ResAtt that combines CNNs,  bidirectional gated recurrent neural network (BiGRU) and attention to enable HAR. Luo et al. \cite{luo2022activity} present an algorithm consisting of three CNN layers and three squeeze-and-excitation (SE) blocks. The SE block serves to adjust the weight of different channels by modeling channelwise relationships.

Several methods achieve multi-scale feature extraction by using convolution operations with different receptive fields. Tang et al. \cite{tang2022multiscale} propose a component named HS block that learns multiscale features for HAR without increasing memory and computational costs. The input features are segmented into distinct groups by this module, and each group undergoes a convolution operation and then directly links with the next layer. Challa et al. \cite{2021AMultibranch} propose a CNN and LSTM hybrid model that feeds signals into three branches composed by CNNs with different kernel sizes. Then two Bi-LSTM modules are stacked to learn temporal features. %Liu et al. \cite{LIU2023_TransTM} proposed a transformer-based model called TransTM that learns multiscale features by convolution layer with different kernel sizes. 
Xu et al. \cite{xu2023multi-channelDCT} propose a novel component that extracts multi-frequency components based on the discrete cosine transform to calculate channel attention factor, and achieves good results. It is also shown that for different activities, the contribution of multi-frequency components is different. For activities such as opening a door or a refrigerator, the high-frequency component contributes more to the model prediction, while the low-frequency component contributed more to quiet activities such as sitting and lying down.

In contrast to the related work covered above, we propose a multilevel heterogeneous model based on wavelet transform for HAR. The wavelet transform can decompose the signal from time and frequency domains without information loss \cite{wang2018multilevel}. It can separate the noise from the signal, thus reducing the impact of noise on subsequent model training. With the deepening of the wavelet transform layer, the sub-component dimension and information decrease. A heterogeneous feature learner is used to learn coarse-to-fine features for multilevel components. Based on the complementarity between multilevel features, we design a cross aggregation module that fuses features to help the model learn more powerful features.

\section{METHODS}
\label{method}
In this section, we describe the proposed model \model~for sensor-based human activity recognition. Initially, we provide a comprehensive overview of the model, followed by an in-depth exposition of each individual module.

\begin{figure*}[t!]
\centering
\includegraphics[width=7in]{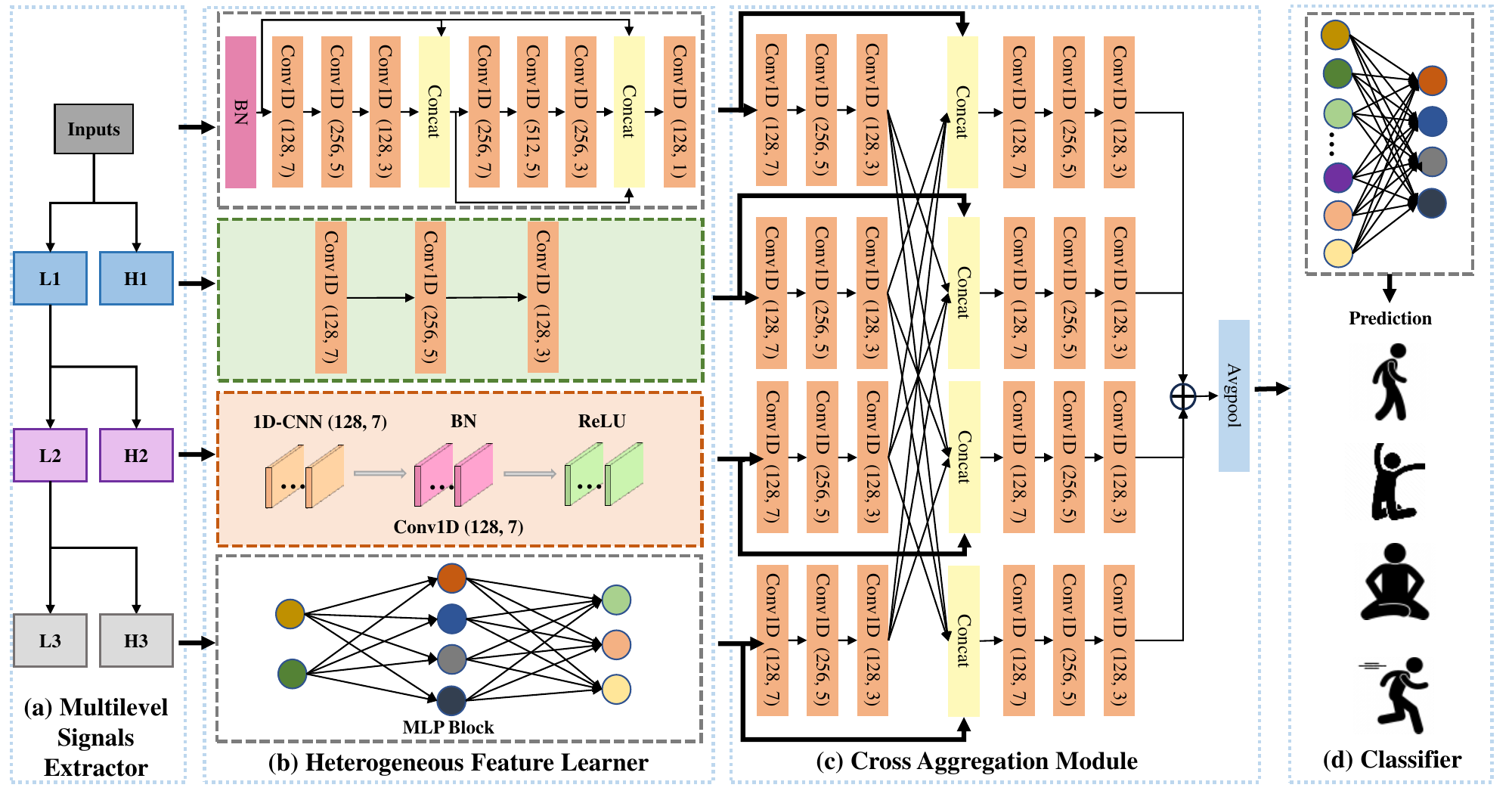}
\caption{The architecture of proposed model. (a) The multilevel signals extractor based on MDWD. (b) Heterogeneous feature learner extracts representational features from multilevel signals. (c) Cross aggregation module integrates multilevel features. (d) A classifier outputs the probability of each human activity according to the integrated features.}
\label{whole_archi}
\end{figure*}

\subsection{Overview}
The architecture of proposed \model, shown in Fig. \ref{whole_archi_ab}, is specifically designed for sensor-based human activity recognition. It contains three main modules: data collection, data preprocessing, deep learning classification model. We collect sensor signals from wearable devices. These signals contain the information when subjects performs various activities, including walking, jogging, sitting, standing, upstairs, and downstairs. This data is often affected by noise and missing values, which introduces significant fluctuations, especially in the high-frequency components. To address these challenges, we employ discrete wavelet decomposition, a robust tool for multi-resolution analysis. In the wavelet domain, the signal demonstrates coherence, with concentrated "energy" primarily present in a few high-amplitude coefficients. Conversely, irrelevant noise is depicted by numerous coefficients characterized by small magnitudes \cite{srivastava2016new}. Thereby wavelet transform can isolate noise from signal coefficients. Fig. \ref{wavelet_visual} provides a visual representation of the noise suppression capabilities of discrete wavelet decomposition in sensor signals. For illustration, we apply MDWD to the Opportunity dataset. This dataset encompasses 45 signals, each of which has a length of 24 units. After decomposition, we acquire both high-frequency (H) and low-frequency (L) sub-signals. In accordance with the tenets of wavelet decomposition, the time resolution for each decomposed component is halved. As a result, the structural and quantitative attributes of the components at each layer, obtained through MDWD, exhibit considerable variation.

Discrete wavelet decomposition serves a dual purpose: it not only suppresses noise but also facilitates the extraction of features across multiple scales. This is essential for capturing the temporal characteristics of human activities that vary significantly across different types of activities and demographic groups. For example, elderly individuals tend to walk at a slower pace with shorter strides, while younger individuals exhibit brisker walking styles with longer strides. To capture these nuances, we design the neural network structure as shown in Fig. \ref{whole_archi}. This structure consists of four modules: multilevel signals extractor, heterogeneous feature learner, cross aggregation module and classifier. We employ heterogeneous feature extractors specifically tailored for multilevel components. These feature extractors are designed with varying levels of complexity to match the intricacies of the high-resolution components and the summarization needs of the low-resolution components. We choose discrete wavelet decomposition to learn multi-scale features for several compelling reasons. First, discrete wavelet decomposition effectively preserves both frequency domain and time domain information without information loss. Second, discrete wavelet decomposition can separate signals of different frequency bands. Due to the high-frequency characteristic of noise, discrete wavelet decomposition can isolate noise from original signals. The heterogeneous feature learner is composed of 1-dimensional convolutional neural networks (1D-CNNs) and multilayer perceptron (MLP). 1D-CNN is useful for the feature extraction of time series data, in which the features of interest are learned through a 1D convolution kernel on each channel signal. Compared with RNN, it is faster and works well. The outputs of heterogeneous feature learner are fed into the cross aggregation module. The multilevel features are concatenated and then processed by three stacked 1D-CNNs to learn fused features. The cross-level features generated from the feature integration module are input into a linear classifier for activity classification. We implement this classifier using a fully connected layer with \(k\) nodes, where \(k\) represents the number of activity categories under consideration. The classifier serves as the final decision-making entity, taking the processed features and mapping them to specific activity categories.

\subsection{Multilevel Signals Extractor}
The architecture of the Multilevel Signals Extractor is illustrated in Fig. \ref{whole_archi}(a). It employs a MDWD to extract multi-resolution components from the raw signal. MDWD is particularly advantageous for this application as it allows for the simultaneous analysis of signal features across multiple scales, thereby capturing both global and local characteristics of the data. Specifically, a series of low-pass and high-pass filters are applied to partition the signal into its low-frequency and high-frequency constituents. The choice of filters is guided by their ability to isolate specific frequency bands, thereby enabling a more nuanced analysis of the signal's characteristics. This decomposition is recursively applied to the low-frequency component obtained from the preceding layer.

MDWD operates in both the time and frequency domains. As the decomposition depth increases, the frequency resolution of the decomposed components improves, while the time resolution diminishes. In each layer, the low-frequency signal encapsulates approximate statistics like the mean and median, whereas the high-frequency signal captures detailed variations, including noise predominantly present in the high-frequency band.

Let \(X = \{X^1, X^2, \ldots, X^C\}\) denote the input signal, where \(C\) is the number of channels. For each channel \(c\), the signal \(X^c = [x^c_1, x^c_2, \ldots, x^c_T]\) undergoes MDWD to extract frequency sub-bands. The decomposition at the \(i\)-th layer employs low-pass filters \(\phi = [\phi_1, \phi_2, \ldots, \phi_K]\) and high-pass filters \(\varphi = [\varphi_1, \varphi_2, \ldots, \varphi_K]\), where \(K \ll T\), to transform the low-frequency component \(L^c(i-1)\) from the previous layer as follows: 
\begin{align}
L^c_n(i) &= \sum_{k=1}^{K} L^c_{n+k-1}(i-1) \phi[k], \\
H^c_n(i) &= \sum_{k=1}^{K} L^c_{n+k-1}(i-1) \varphi[k].
\end{align}
The complete set of components from MDWD is denoted as \(W^c = \{H^c(1), H^c(2), \ldots, H^c(I), L^c(I)\}\), where \(I\) is the number of decomposition layers. These components are then aggregated into a two-dimensional matrix \(W\), defined as
\[
W = [X, H(1), H(2), \ldots, H(I), L(I)],
\]
where \(H(i) = [H^1(i), H^2(i), \ldots, H^C(i)]\) and\(L(I) = [L^1(I), L^2(I), \ldots, L^C(I)]\) are matrices formed by the high- and low-frequency components across all channels, respectively.

As the depth of wavelet decomposition increases, the temporal resolution of sub-signal components progressively decreases, while the frequency resolution steadily increases. When the decomposition depth is excessively high, the deep-level components contain insufficient information, rendering the model unable to capture meaningful features. Conversely, if the decomposition depth is too shallow, there may be an issue of inadequate feature extraction. To choose the best decomposition layer, we contrast the impact of different levels of wavelet decomposition on model performance in sensitivity analysis experiments.

\subsection{Heterogeneous Feature Learner}
% % 先解释为什么要做异构，然后介绍异构的结构
% Wavelet decomposition is a reversible operation, and the original signal can be regenerated according to the components and filters after decomposition. With the deepening of decomposition layers, the dimension of the component decreases and the information contained in it decreases. If the same deep learning structure is used to extract features for each layer component, it will appear that the deep component is fully learned while the upper component is underfitted. Therefore, we design different feature extractors for each layer component.

As illustrated in Fig. \ref{whole_archi}(b), the complexity of the sub-feature extractor is inversely proportional to the depth of the corresponding layer. Each sub-feature extractor is tailored to the nature of the decomposed signal it processes, as detailed below.

For \(H3\), the high-frequency component obtained after three levels of wavelet decomposition, a neural network comprising three fully connected layers with 128 nodes each is employed. Each of these layers is succeeded by a LeakyReLU activation function and a dropout mechanism. LeakyReLU is designed to mitigate the vanishing gradient problem and reduce the risk of overfitting. Mathematically, it is defined as \(y = \max(0, x) + l \times \min(0, x)\), where \(l\) is a small positive constant. A basic Multilayer Perceptron (MLP) module is expressed as:
\begin{equation}
\label{mlp}
z_i = \text{LeakyReLU}(W_iz_{i-1} + b_i),
\end{equation}
where \(z_i\) is the output of the \(i\)-th MLP layer, and \(W_i\) and \(b_i\) are the corresponding parameters.

For \(H2\), a 1D-CNN block serves as the feature extractor. This block consists of a 1D-CNN with a kernel size of 7, followed by batch normalization (BN) and a ReLU activation function. The architecture is termed a Conv1D block, a nomenclature we adopt throughout this paper. The 1D-CNN captures local features, while BN enhances model convergence and generalization by normalizing the input distribution. %A element of CNN mathematical formulation  is:
% \begin{equation}
% \label{1d-cnn}
% o_{i, j} = \sum_{k=1}^{K} W_{k} x^c_{i+k-1, j},
% \end{equation}
% and the Conv1D block is computed as:
% \begin{equation}
% \label{cnn-block}
% z = \text{ReLU}(\text{BN}(O)),
% \end{equation}
% where \(O\) is the output matrix of the CNN.

For \(H1\), three Conv1D blocks with kernel sizes of 7, 5, and 3 are employed. As the network depth increases, the receptive field expands, enabling the extraction of more complex features. 

For the original signals, which contain a mix of valuable information and noise, a more intricate architecture comprising seven Conv1D blocks is utilized. We introduce residual connections \cite{he2016deep} between these blocks to facilitate the learning of residual features, thereby ameliorating the vanishing gradient problem and enhancing the model's generalization capabilities.

\subsection{Cross Aggregation Module}
Traditional approaches of feature fusion mainly employ simple operations such as concatenation, addition, and multiplication. While computationally efficient, these linear aggregation techniques risk the loss of valuable information by failing to selectively emphasize specific features~\cite{dai2021attentional}.

To capture fine-grained details at shallow layers and high-level summaries at deeper ones, we present a novel cross aggregation module. This module harmoniously fuses features from multiple levels, harnessing their inherent strengths and relationships, as depicted in Figure \ref{whole_archi}(c).
%Recognizing intricate features at shallower layers capture fine-grained details and those at deeper layers offer a high-level summary, we introduce a novel cross-aggregation module. This module synergistically combines features across multiple layers, capitalizing on their inherent strengths and interrelationships, as depicted in Figure \ref{whole_archi}(c).

Within this architecture, we deploy three Conv1D blocks to extract auxiliary features from each level features. The kernel sizes for these blocks are judiciously chosen as 7, 5, and 3, while maintaining a uniform filter count of 128 across all blocks.

To mathematically formalize the cross-aggregation process, let \( \mathcal{F}_i \) denote the features of \( i \)-th level. The auxiliary feature \( \mathcal{A}_i \) extracted from \( \mathcal{F}_i \) can be represented as:
\begin{equation}
\mathcal{A}_i = Conv1D(\mathcal{F}_i).
\end{equation}
The cross-aggregated feature \( \mathcal{H}_i \) for layer \( i \) is then given by:
\begin{equation}
\mathcal{H}_i = Conv1D(\mathcal{A}_1 \oplus \mathcal{A}_2 \oplus \ldots \oplus \mathcal{A}_n \oplus \mathcal{F}_i),
\end{equation}
where \( \oplus \) denotes concatenation. The inputs of \(\mathcal{H}_i\) include \(\mathcal{F}_i\) and all other auxiliary features except \( \mathcal{A}_i \).

Finally, the cross-aggregated features are integrated using bitwise addition to produce the final output, \( \mathcal{O} \), of the module:
\begin{equation}
\mathcal{O} = \sum_{i=0}^{n} \mathcal{H}_i %\plus \mathcal{H}_2 \plus \ldots \plus \mathcal{H}_n.
\end{equation}

The computational complexity of the cross-aggregation module is \( O(n \times m \times k) \), where \( n \) is the number of layers, \( m \) is the number of features per layer, and \( k \) is the kernel size. In contrast, traditional methods that employ simple concatenation or addition have a complexity of \( O(n \times m) \). While the cross-aggregation module incurs a slightly higher computational cost, the enhancement in feature expressiveness and model performance justifies this trade-off.

\subsection{Classifier}

The final stage of our neural network model is dedicated to the task of classifying human activities based on the features extracted from the preceding layers. For this purpose, we employ a softmax classifier, integrated into the neural network, which outputs a \( K \)-dimensional vector of real numbers, where \( K \) represents the number of unique activity labels.

The softmax classifier is particularly advantageous in this neural network architecture for several reasons. First, it allows for end-to-end training of the entire network through backpropagation, thereby ensuring that the features and classifier are optimized jointly. Second, it is computationally efficient, making it well-suited for real-time applications. 

The classifier calculates the probability \( \hat{y}_{i,k} \) for each sample \( i \) belonging to class \( k \) as follows:
\begin{equation}
\label{softmax}
\hat{y}_{i,k} = \frac{e^{o_{i,k}}}{\sum_{k=1}^{K}{e^{o_{i,k}}}},
\end{equation}
where \( o_{i,k} \) is the \( k \)-th dimension of the feature output, generated from the final layer of our model.

For assessing the performance of classifier, the cross-entropy loss function is employed:
\begin{equation}
\label{cross-entropy}
L(y_i, \hat{y}_i) = -\sum_{k=1}^K  y_{i, k} \log(\hat{y}_{i,k}) + (1-y_{i, k}) \log(1-\hat{y}_{i, k}),
\end{equation}
where \( y_{i} \) is the true label for sample \( i \) in class \( k \), taking binary values of either 0 or 1.

By opting for a softmax classifier within this neural network architecture, we aim to achieve a harmonious balance between computational efficiency and predictive accuracy, thereby optimizing the model's performance.

\section{EXPERIMENT}
\label{experiment}
%In this section, we apply our proposed model to seven publicly available Human Activity Recognition (HAR) datasets, namely UCI-HAR, WISDM, PAMAP2, Opportunity, RealWorld, Realdisp, and Skoda. Detailed descriptions of these datasets are provided in Part A. We introduce the evaluation metrics in Part B. In Part C, we conduct a comparative analysis, pitting our model, referred to as \model, against other state-of-the-art methods. Part D is dedicated to assessing the robustness of our model in handling missing values, which we validate through a leave-sensor-out experiment. Finally, in Part E, we evaluate the significance of each module through an ablation analysis.
In this section, we conduct experiments to evaluate our proposed model. 

\subsection{Implementation and experimental settings}
 We implement our model with PyTorch v1.8.0, a widely-used deep learning framework. All experiments are conducted on a server equipped with an Intel(R) Xeon(R) Silver 4216 CPU (2.10 GHz) and 500GB of memory. To accelerate training and evaluation processes, we employ a GeForce RTX 2080 Ti GPU. For the training phase, we set the learning rate of the Adam optimizer to 5e-4 and set batch size to 128. We split the dataset into a training set and a test set. Following each training iteration, we evaluate model's performance on the test set. Early stopping optimization mechanism is applied during the training progress. In our experiments, we opt for the Haar function as the fundamental wavelet for our wavelet decomposition due to its simplicity in structure, ease of interpretation, and rapid computational speed.

\subsection{Datasets and preprocessing}

Our experimental evaluation leverages seven diverse datasets, each offering unique challenges and opportunities for HAR. Below, we provide a detailed description of each dataset and the preprocessing steps undertaken to ensure uniformity across experiments.
\begin{itemize}[left=0pt]
\item \textbf{WISDM:} The Wireless Sensor Data Mining (WISDM) dataset~\cite{2013Activity} serves as a well-established benchmark in HAR research. It comprises three-axial accelerometer signals from Android smartphones, sampled at 20 Hz. Data are collected from 29 volunteers performing six distinct activities. The signals are segmented into 10-second windows, each containing 200 readings.

\item \textbf{UCI-HAR:} Curated by Bulbul et al.~\cite{2019A}, this dataset involves 30 subjects performing six activities while wearing smartphones on their waists. The dataset includes accelerometer and gyroscope readings at a 50 Hz sampling rate.

\item \textbf{PAMAP2:} This dataset~\cite{reiss2012introducing} is designed to monitor daily living activities using three inertial measurement units. After preprocessing, it comprises 5,333 segments, each containing 600 observations at a 100 Hz sampling rate.

\item \textbf{Opportunity:} This dataset~\cite{roggen2010collecting} is collected in a breakfast scenario and includes readings from various types of sensors. Our focus is on the data collected from five body-worn inertial measurement units (IMUs).

\item \textbf{RealWorld:} Described in~\cite{sztyler2016body}, this dataset collects activities from 15 users. We use data from 13 users, each performing eight different physical activities.

\item \textbf{Realdisp:} This dataset~\cite{banos2012benchmark} is curated to explore the effects of sensor displacements. We limit our analysis to data from 10 users performing 33 distinct fitness activities.

\item \textbf{Skoda:} Detailed in~\cite{stiefmeier2008wearable}, this dataset captures 10 distinct manipulative gestures performed by an assembly-line worker. We exclusively use data from the right arm.
\end{itemize}

To maintain consistency, we apply a common preprocessing method~\cite{liu2020giobalfusion}. For the Opportunity, RealWorld, and Realdisp datasets, we restrict our analysis to accelerometer, gyroscope, and magnetometer data, resulting in nine raw features per sensor. For the Skoda dataset, we only use the 3-axis acceleration data, yielding three raw features. The sampling rates are adjusted as needed to align with the literature.

By employing these datasets and preprocessing steps, we conduct a thorough evaluation of our model, thereby ensuring its robustness and generalizability across diverse scenarios.

\subsection{Evaluation Metrics}

To access the model \model, we employ a comprehensive set of evaluation metrics, described below:

\begin{itemize}[left=0pt]
\item \textbf{Accuracy}: The ratio of correctly classified samples to total samples, accuracy is mathematically expressed as:
\begin{equation}
\label{eq:accuracy}
\text{Accuracy} = \frac{TP + TN}{TP + FP + TN + FN},
\end{equation}
where \(TP\), \(FP\), \(TN\), and \(FN\) denote the number of true positive, false positive, true negative, and false negative samples, respectively.

\item \textbf{Recall}: The proportion of actual positive samples that are correctly identified to predicted positive samples, and is given by:
\begin{equation}
\label{eq:recall}
\text{Recall} = \frac{TP}{TP + FN}.
\end{equation}

\item \textbf{Precision}: The proportion of predicted positive samples that are actually positive to the actual positive samples, as described by:
\begin{equation}
\label{eq:precision}
\text{Precision} = \frac{TP}{TP + FP}.
\end{equation}

\item \textbf{F1 Score}: The harmonic mean of recall and precision, it provides a balanced measure of a model's performance, particularly for class-imbalanced tasks. It is formulated as:
\begin{equation}
\label{eq:f1_score}
\text{F1 Score} = \frac{2 \times \text{Recall} \times \text{Precision}}{\text{Recall} + \text{Precision}}.
\end{equation}
\end{itemize}

In comparative analyses with state-of-the-art (SOTA) models, the four metrics are utilized in sections discussing the influence of noise and ablation analysis.

\subsection{Baseline methods}
We compare our proposed method with the following 28 baseline methods for human activity recognition:
% \ask{Please find all the citaions for the baseline methods}
\begin{itemize}[left=0pt]
\item \textbf{DeepConvLSTM} \cite{ordonez2016deep}: A deep learning model that combines CNNs and LSTMs to learn spatial and temporal features from raw sensor data.
\item \textbf{Att. Model} \cite{murahari2018attention}: A deep learning model that uses an attention mechanism to selectively focus on the most relevant features for each activity class.
\item \textbf{GraphConvLSTM} \cite{han2019graphconvlstm}: A deep learning model that combines graph convolutional networks (GCNs) and LSTM networks to model the spatial relationship and temporal dependency of sensor signals.
% \item \textbf{Attend and Discriminate} \cite{abedin2021attend}: A semi-supervised learning framework that uses a spanning forest algorithm and a silhouette-based filtering method to augment the labeled data and train a CNN-LSTM model for activity recognition.
\item \textbf{Attend and Discriminate} \cite{abedin2021attend}: A deep learning framework that uses a mixup method to augment the labeled data and train a CNN-GRU model for activity recognition.
\item \textbf{HAR-PBD} \cite{wang2021leveraging}: A hierarchical deep learning model that integrates HAR with protective behavior detection (PBD) using graph-convolution and LSTM networks and a class-balanced focal categorical-cross-entropy loss function.
% \item \textbf{DynamicWHAR} \cite{miao2022towards}: A dynamic weighted hybrid attention recurrent network that incorporates a weighted fusion of CNNs and LSTM networks with a dynamic attention mechanism to capture the discriminative features of human activities.
\item \textbf{DynamicWHAR} \cite{miao2022towards}: a GCN-based dynamic inter-sensor correlations learning framework that extract feature using 1D-CNN and then achieve the information interaction between multiple sensors by GCN.
\item \textbf{SSRCAN} \cite{2020A}: A self-supervised representation learning method that uses contrastive attention networks to learn high-level representations from unlabeled sensor data and fine-tunes them withv small labeled samples for activity recognition.
% \item \textbf{AIA} \cite{8727452}: An adaptive incremental aggregation method that uses a sliding window approach to segment the sensor data and aggregates the features extracted by a CNN-LSTM model using an adaptive weighting scheme based on the confidence scores of the predictions.
\item \textbf{AIA} \cite{8727452}: A HAR system that collect data from wearable sensor and train a CNN model to classify activities.
\item \textbf{LabelForest} \cite{ma2019labelforest}: A non-parametric semi-supervised learning framework that uses a spanning forest algorithm and a silhouette-based filtering method to infer labels for unlabeled data and train a random forest classifier for activity recognition.
\item \textbf{LSTM-CNN} \cite{xia2020lstm}: A deep learning model combining convolutional layers with LSTM for automatic extraction of activity features from raw data collected by mobile sensors.
\item \textbf{SSFT} \cite{saeed2019multi}: A semi-supervised feature transfer method that uses an autoencoder to learn low-dimensional representations from unlabeled data and transfers them to a supervised classifier trained on labeled data for activity recognition.
\item \textbf{KNN-LS-SVM} \cite{amer2020localized}: A hybrid method that combines k-nearest neighbors (KNN) and least squares support vector machines (LS-SVM) to classify human activities based on handcrafted features extracted from sensor data.
\item \textbf{MHCA} \cite{zhang2019novel}: A multi-head convolutional attention network that uses parallel convolution structure and an attention mechanism to learn multi-scale features from sensor data for activity recognition.
% \item \textbf{DanHAR} \cite{gao2021danhar}: A domain adaptation network that uses an adversarial learning strategy to align the feature distributions of different domains and reduce the domain discrepancy for cross-domain activity recognition.
\item \textbf{DanHAR} \cite{gao2021danhar}: A dual attention framework that integrates channel attention and temporal attention within a CNN, aimed at addressing spatial-temporal dependencies in multimodal sensing signals.
% \item \textbf{DARMTA} \cite{alsheikh2015deep}: A domain adaptation recurrent multitask attention network that uses a shared LSTM encoder, multiple task-specific decoders, and an adversarial discriminator to learn domain-invariant and task-specific features for cross-domain multitask activity recognition.
% \item \textbf{DARMTA} \cite{alsheikh2015deep}: a hybrid approach of deep learning and hidden Markov models that integrates the hierarchical representations deep activity recognition models with the stochastic modeling of temporal sequences in the hidden Markov models.
% \item \textbf{DSSL} \cite{qian2019distribution}: A deep semi-supervised learning method that uses a ladder network to jointly optimize the supervised and unsupervised losses and learn robust representations from labeled and unlabeled data for activity recognition.
\item \textbf{DSSL} \cite{qian2019distribution}: A distribution-based semi-supervised learning method that maps all data to reproducing kernel Hilbert space (RKHS) and alter the RKHS according to underlying geometry structure of the unlabeled distributions and train a classifier using labeled data.
\item \textbf{Lego-CNN} \cite{tang2020efficient}: A modular convolutional neural network that consists of multiple interchangeable building blocks with different configurations and learns the optimal architecture for activity recognition using reinforcement learning.
\item \textbf{MVAN} \cite{2019Hierarchical}: A multi-view attention network that uses multiple CNNs with different input views (time, frequency, time-frequency) and an attention mechanism to fuse the multi-view features for activity recognition.
\item \textbf{M-U-Net} \cite{2019Human}: A multi-scale U-Net that uses multiple U-shaped convolutional blocks with different scales and skip connections to learn hierarchical features from sensor data for activity recognition.
\item \textbf{OCL} \cite{2019Online}: An online contrastive learning method that uses a memory bank to store the representations of historical samples and learns contrastive features from online batches of sensor data for activity recognition.
\item \textbf{LWTCNN} \cite{2020The}: A lightweight temporal convolutional neural network that uses depthwise separable convolutions and residual connections to reduce the model complexity and learn temporal features from sensor data for activity recognition.
% \item \textbf{SRRS} \cite{2019Wearable}: A sparse recurrent residual stack that uses a stack of residual LSTM blocks with sparse connections and dropout regularization to learn long-term dependencies from sensor data for activity recognition.
\item \textbf{SRRS} \cite{2019Wearable}: A discriminant approach that blends S transform and supervised regularization-based robust subspace (SRRS) learning method to extract robust feature for sensor-based activity recognition.
\item \textbf{GRU-D} \cite{che2018recurrent}: A recurrent neural network that uses gated recurrent units (GRUs) to handle missing values and irregular time intervals in multivariate time series data. GRU-D incorporates two representations of missing patterns: masking and time interval, and learns to impute missing values dynamically during the forward pass. 
\item \textbf{SeFT} \cite{horn2020set}: A spiritual emotional freedom technique (SeFT) that combines tapping on specific acupressure points with positive affirmations to reduce depression and anxiety for chronic renal failure patients undergoing hemodialysis. SeFT consists of three stages: setup, tune in, and tapping, and aims to neutralize the negative energy and emotions in the body. 
\item \textbf{mTAND} \cite{shukla2021multi}: A model named MultiTime Attention Networks that utilizes an attention mechanism to create a fixed-length representation of time series. mTAND can handle data with varying time intervals, and can be used for interpolation and classification tasks. 
% \item \textbf{IP-Net} \cite{shukla2019interpolation}: A graph-guided network that predicts interaction points, which directly localize and classify the human-object interactions in an image. IP-Net uses a graph learning module to extract the uni-directed relations among variables, a mix-hop propagation layer to capture the spatial dependencies, and a dilated inception layer to capture the temporal dependencies. 
\item \textbf{IP-Net} \cite{shukla2019interpolation}: a deep learning architecture that involves a semi-parametric interpolation network, facilitating information sharing across dimensions, followed by a prediction network. 
\item \textbf{DGM2-O} \cite{wu2021dynamic}: A generative model that tracks the transition of latent clusters rather than isolated feature representations. DGM2-O uses an ordinary differential equation (ODE) solver to model the transition dynamics of the latent variables, and a dynamic Gaussian mixture distribution to model the observation likelihood. DGM2-O can handle missing values, variable selection, and uncertainty quantification. 
% \item \textbf{MTGNN} \cite{wu2020connecting}: A multi-time graph neural network for metro passenger flow forecasting that incorporates both short-term and long-term historical information. MTGNN uses a graph convolutional network (GCN) to capture the spatial correlations among metro stations, and a gated recurrent unit (GRU) to capture the temporal dependencies among passenger flows. 
\item \textbf{MTGNN} \cite{wu2020connecting}: A graph neural network model which combines graph with mix-hop propagation to extract latent spatial dependencies between variables.%An end-to-end framework for multivariate time series forecast that composed by a graph learning, graph convolution, and temporal convolution modules. MTGNN uses mix-hop propagation layer and a dilated inception layer to capture spatial and temporal dependencies within the time series.
\item \textbf{RAINDROP} \cite{zhang2021graph}: A graph neural network that constructs graph for each sample and utilizes message passing operator to model time-varying dependencies between sensors. RAINDROP can be used for classification and interpretation of temporal dynamics.%embeds irregularly sampled and multivariate time series while also learning the dynamics of sensors purely from observational data. RAINDROP represents every sample as a separate sensor graph and models time-varying dependencies between sensors with a novel message passing operator. RAINDROP can be used for classification and interpretation of temporal dynamics.
\end{itemize}

\begin{table*}[t!]
\caption{Comparisons with the superior models for Opportunity, RealWorld, Reasdisp, and Skoda datasets}
\label{tab:four_datasets}
\centering
\begin{tabular}{l|l|cc|cc|cc|cc}
\hline
\multirow{2}{*}{Model}  & \multirow{2}{*}{Year} & \multicolumn{2}{c|}{Opportunity}      & \multicolumn{2}{c|}{RealWorld}        & \multicolumn{2}{c|}{Realdisp}         & \multicolumn{2}{c}{Skoda}            \\ \cline{3-10} 
                        &                       & \multicolumn{1}{c|}{Accuracy} & F1    & \multicolumn{1}{c|}{Accuracy} & F1    & \multicolumn{1}{c|}{Accuracy} & F1    & \multicolumn{1}{c|}{Accuracy} & F1    \\ \hline
                        
DeepConvLSTM            & 2018                  & \multicolumn{1}{c|}{69.30}     & 61.32 & \multicolumn{1}{c|}{72.42}    & 62.35 & \multicolumn{1}{c|}{85.61}    & 83.56 & \multicolumn{1}{c|}{90.31}    & 88.63 \\ \hline
Att. Model              & 2018                  & \multicolumn{1}{c|}{71.64}    & 63.43 & \multicolumn{1}{c|}{73.84}    & 65.90 & \multicolumn{1}{c|}{88.37}    & 87.70 & \multicolumn{1}{c|}{91.55}    & 90.76 \\ \hline
GraphConvLSTM           & 2019                  & \multicolumn{1}{c|}{70.38}    & 62.17 & \multicolumn{1}{c|}{72.71}    & 63.56 & \multicolumn{1}{c|}{89.94}    & 89.04 & \multicolumn{1}{c|}{91.19}    & 89.88 \\ \hline
Attend and Discriminate & 2021                  & \multicolumn{1}{c|}{72.75}    & 64.18 & \multicolumn{1}{c|}{74.58}    & 72.69 & \multicolumn{1}{c|}{90.46}    & 89.76 & \multicolumn{1}{c|}{92.16}    & 90.87 \\ \hline
HAR-PBD                 & 2021                  & \multicolumn{1}{c|}{71.82}    & 62.37 & \multicolumn{1}{c|}{74.04}    & 71.73 & \multicolumn{1}{c|}{91.41}    & 90.48 & \multicolumn{1}{c|}{88.37}    & 86.26 \\ \hline
DynamicWHAR             & 2022                  & \multicolumn{1}{c|}{74.27}    & 66.13 & \multicolumn{1}{c|}{\textbf{76.63}}    & \textbf{75.13} & \multicolumn{1}{c|}{92.58}    & 91.93 & \multicolumn{1}{c|}{93.80}    & 91.26 \\ \hline
Ours                    & 2023                  & \multicolumn{1}{c|}{\textbf{80.00}}    & \textbf{73.59} & \multicolumn{1}{c|}{75.87}    & 69.13 & \multicolumn{1}{c|}{\textbf{92.89}}    & \textbf{93.24} & \multicolumn{1}{c|}{\textbf{96.60}}    & \textbf{95.63} \\ \hline
\end{tabular}
\end{table*}

\begin{table}[t!]
\caption{ Comparisons with the SOTA models for UCI-HAR datasets}
\label{tab:UCI-HAR}
\centering
\begin{tabular}{l|c|c|c|c }
\hline
Model                & Accuracy & Precision & Recall & F1    \\ \hline
SSRCAN      & 83.12    & 71.98     & 72.26 & 72.12 \\ \hline
AIA        & 87.65    & 88.23     & 88.21 & 88.22 \\ \hline
LabelForest  & 91.23    & 91.34     & 91.21 & 91.27 \\ \hline
LSTM-CNN     & 90.34    & 91.34     & 91.32 & 91.33 \\ \hline
SSFT       & 91.23    & 90.97     & 90.93 & 90.65 \\ \hline
Ours                 & \textbf{95.05}  & \textbf{95.40} & \textbf{95.00}   & \textbf{95.14} \\
\hline
\end{tabular}
\end{table}

\begin{table}[t!]
\caption{ Comparisons with the SOTA models for WISDM datasets}
\label{tab:WISDM}
\centering
\begin{tabular}{l|c|c|c|c}
\hline
Models                 & Accuracy & Precision & Recall & F1   \\ \hline
KNN-LS-SVM  & 95.40    & 95.00     & 96.00  & 95.50  \\ \hline
MHCA         & 95.50    & 95.60     & 95.20  & 95.40  \\ \hline
DanHAR         & 98.85    & 98.34     & 98.35  & 98.34 \\ \hline
%DARMTA        & 98.23    & 98.32     & 98.34  & 98.33 \\ \hline
DSSL          & 56.70    & 56.70     & 56.30  & 56.50
\\ \hline
Lego-CNN       & 97.51    & 97.60     & 97.56  & 97.58 \\ \hline
MVAN          & 93.10    & 92.90     & 93.10  & 93.00    \\ \hline
M-U-Net      & 96.40    & 96.20     & 96.40  & 96.30  \\ \hline
SSFT          & 90.01    & 89.99     & 85.68  & 86.86 \\ \hline
OCL          & 81.30    & 82.32     & 81.23  & 81.77 \\ \hline
LWTCNN       & 98.82    & 98.79     & 98.83  & 98.81 \\ \hline
SRRS         & 93.50    & 93.40     & 92.60  & 92.30  \\ \hline
Ours                   & \textbf{99.33} & \textbf{98.97} & \textbf{99.24} & \textbf{99.10}  \\ \hline
\end{tabular}
%}
\end{table}
 
\begin{table}[t!]
\caption{ Comparisons with the SOTA models for PAMAP2 datasets}
\label{tab:PAMAP2}
\centering
\begin{tabular}{l|c|c|c|c}
\hline 
Models                 & Accuracy & Precision & Recall & F1    \\ \hline
Transformer & 83.5 & 84.8  & 86.0  & 85.0  \\ \hline
Trans-mean &	83.7 &	84.9 &	86.4 &	85.1 \\ \hline
GRU-D & 83.3 & 84.6  & 85.2 & 84.8  \\ \hline
SeFT & 67.1  & 70.0  & 68.2  & 68.5  \\ \hline
mTAND & 74.6  & 74.3  & 79.5  & 76.8 \\ \hline 
IP-Net & 74.3  & 75.6  & 77.9  & 76.6 \\ \hline
DGM$^2$-O & 82.4  &  85.2 & 83.9  & 84.3  \\ \hline
MTGNN & 83.4  & 85.2  & 86.1  & 85.9 \\ \hline
RAINDROP & 88.5  & 89.9  & 89.9  & 89.8 \\ \hline
Ours                   & \textbf{95.88} & \textbf{96.72} & \textbf{96.41} & \textbf{96.50} \\ \hline
\end{tabular}
%}
\end{table}
\subsection{Comparison of SOTA Approaches}
In this subsection, a comparative analysis is conducted between our proposed model and 28 existing models within the domain of HAR. The datasets employed for this analysis include Opportunity, RealWorld, Realdisp, Skoda, UCI-HAR, WISDM, and PAMAP2. Our evaluation relies on experimental outcomes previously reported in the literature that introduced these benchmark models. Specifically, we assess the efficacy of our model by examining its accuracy and F1-score metrics. This enables us to substantiate its robustness, generalizability, and superiority. Moreover, the datasets offer diverse contexts, characterized by variations in sensor types, sensor placement, types of activities, and data quality. In our analysis, we include 28 baseline methods, each evaluated on distinct subsets of the aforementioned seven datasets. The comparative results are presented in Table \ref{tab:four_datasets}, Table \ref{tab:UCI-HAR}, Table \ref{tab:WISDM}, and Table \ref{tab:PAMAP2}.
% \COMMENT{Maybe we can delete the experiment result on RealWorld dataset. Our model does not achieve the best performance on RealWorld dataset. And on noise experiment, ablation analysis and sensitivity analysis, this dataset is not considered. }

Table \ref{tab:four_datasets} presents the comparison result of our model and other six models on four datasets, namely Opportunity, RealWorld, Realdisp, and Skoda. our model has the top performance with accuracy at 86.34\% and F1 score at 82.90\%. The second-best model, DynamicWHAR, has an accuracy of 84.32\% and an average F1 score of 81.14\%.  The accuracy and F1 score of the least performing model, DeepConvLSTM, are 6.93\% and 9.11\% lower than our model. These results demonstrate that our model can effectively learn from the multimodal sensor data and distinguish between various activities in different scenarios.

The comparison result on UCI-HAR dataset is shown in Table \ref{tab:UCI-HAR}. While SSRCAN exhibits competitive accuracy at 83.12\%, AIA demonstrates an improvement, achieving an accuracy of 87.65\%. Further advancements are observed with LabelForest, which outperforms AIA with an accuracy of 91.23\%. LSTM-CNN maintains a consistently high accuracy of 90.34\%. The SSFT model's accuracy is on par with other models, at 91.23\%. However, our developed model surpasses all competitors, attaining a remarkable accuracy of 95.05\%. This substantial quantitative leap signifies the efficacy of our model in capturing intricate patterns within the UCI-HAR datasets, establishing it as the leading performer among the compared models.

As depicted in Table \ref{tab:WISDM}, our model also outperforms all other models on the WISDM dataset. KNN-LS-SVM and MHCA showcase commendable accuracy at 95.40\% and 95.50\%, respectively, demonstrating their effectiveness in activity recognition. Notably, DanHAR exhibits remarkable proficiency, achieving an outstanding accuracy of 98.85\%, making it a standout performer in the comparison. Lego-CNN also delivers high accuracy at 97.51\%, further emphasizing the competitiveness of convolutional neural network architectures in this context. Our proposed model surpasses all counterparts, setting a new benchmark with an accuracy of 99.33\%. This substantial improvement substantiates the efficacy of our approach in capturing nuanced patterns within the WISDM datasets, surpassing the state-of-the-art models by a considerable margin.

As shown in Table \ref{tab:PAMAP2}, the PAMAP2 dataset showcases the superior performance of our model, achieving the highest accuracy. Transformer, Trans-mean, and GRU-D display competitive accuracy levels at around 83-83.7\%, demonstrating their effectiveness in capturing complex patterns within the sensor data. Notably, SeFT and mTAND exhibit lower accuracy, indicating potential limitations in their approach. $DGM^2-O$ and MTGNN showcase commendable accuracy, aligning with the state-of-the-art models in the field. RAINDROP achieves a notable accuracy of 88.5\%, showcasing its proficiency in activity recognition. However, our proposed model surpasses all competitors, achieving an outstanding accuracy of 95.88\%. This substantial improvement solidifies the standing of our model as a leader in the domain of activity recognition based on the PAMAP2 datasets.

In summary, the experimental results demonstrate the superiority of \model over the existing remarkable models across seven widely used human activity recognition datasets. The effectiveness of \model can be attributed to its robust learning capability and the extraction of abundant features from the raw sensor signals.

\subsection{Robust To Missing Values}
In order to evaluate model performance when missing values exist, we conduct two experiments: leave-fixed-sensors-out and leave-random-sensors-out. For the former, all the observations of specific sensors are masked in whole validation set (training samples are complete). For the latter, all observations are randomly hidden as missing values in the validation set, with these missing values replaced by zeros during the validation process. %we select a subset of sensors and regard them as missing by replacing all of their observations with zeros in each validation sample. 
The performance of various machine learning methods is evaluated on the PAMAP2 dataset \cite{reiss2012introducing}. The results are reported in Table \ref{exp_PAM}. The evaluation metrics include accuracy (A), precision (P), recall (R), and F1 score (F1).

The ratio column in Table \ref{exp_PAM} indicates the percentage of sensors that are hidden in each scenario, ranging from 10\% to 50\%. A higher ratio means more missing values and more difficulty for the methods. At a 10\% missing value ratio, our proposed model outperforms other algorithms, achieving the highest accuracy, precision, recall, and F1 scores in both fixed and random sensor out scenarios. This trend persists across higher missing value ratios, where our model consistently demonstrates superior performance. % Notable, in scenarios with 50\% missing values, our model achieves an accuracy of 60.1\% and 66.7\% in fixed and random sensor out cases, respectively, surpassing all other compared algorithms. %The experiment results demonstrate the robustness of our model. For instance, for the 10\% left-out sensor case, our method achieves accuracy at 83.9\% and F1 score at 85.0\% for fixed sensor out. The accuracy and F1 score is higher than the second-best method, RAINDROP, 6.7\% and 9.8\% respectively. Similarly, for random sensor out, our method achieves an accuracy of 89.9\% and an F1 score of 91.1\%, which are also significantly higher than those of RAINDROP, which has an accuracy of 76.7\% and an F1 score of 78.6\%.
Notable, as the ratio increases to 50\%, our method still maintains a high performance for both scenarios, while most of other methods suffer from a sharp decline. For fixed sensor out, our method achieves accuracy at 60.1\% and F1 score at 59.7\%. However, for random sensor out, our method achieves an accuracy of 66.7\% and an F1 score of 67.4\%, which are notably higher than those of RAINDROP.

The experimental results demonstrate that our method is effective and robust in dealing with sensor missing situations in the PAMAP2 dataset, and it can provide a better classification performance than other approaches.

\begin{table}[t!]
\scriptsize
\centering
\caption{Classification performance with SOTA algorithms on the PAMAP2 dataset with missing value. }
\vspace{-3mm}
\label{exp_PAM}
\resizebox{\linewidth}{!}{ 
{\fontsize{30}{32}\selectfont
\begin{tabular}{c|l|l|l|l|l|ll|l|l|l} 
\hline
\multirow{2}{*}{Ratio} & \multicolumn{1}{c}{\multirow{2}{*}{Methods}} & \multicolumn{4}{c}{Fixed sensor out}                                                              &  & \multicolumn{4}{c}{Random sensor out}                                                              \\ 
\cline{3-6}\cline{8-11}
                       & \multicolumn{1}{c}{}                         & A                      & P                      & R                      & F1                     &  & A                      & P                      & R                      & F1                      \\ 
\cline{1-6}\cline{8-11}
\multirow{7}{*}{10\%}  & Transformer                                  & 0.603                  & 0.578                  & 0.598                  & 0.572                  &  & 0.609                  & 0.584                  & 0.591                  & 0.569                   \\
                       & Trans-mean                                   & 0.604                  & 0.618                  & 0.602                  & 0.580                  &  & 0.624                  & 0.596                  & 0.637                  & 0.627                   \\
                       & GRU-D                                        & 0.654                  & 0.726                  & 0.643                  & 0.636                  &  & 0.684                  & 0.742                  & 0.708                  & 0.720                   \\
                       & SeFT                                         & 0.589                  & 0.625                  & 0.596                  & 0.596                  &  & 0.400                  & 0.408                  & 0.410                  & 0.399                   \\
                       & mTAND                                        & 0.588                  & 0.595                  & 0.644                  & 0.618                  &  & 0.534                  & 0.548                  & 0.570                  & 0.559                   \\
                       & RAINDROP                                     & 0.772                  & 0.823                  & 0.784                  & 0.752                  &  & 0.767                  & 0.799                  & 0.779                  & 0.786                   \\
                       & Ours  & \textbf{0.839} & \textbf{0.844} & \textbf{0.876} & \textbf{0.850} &   & \textbf{0.899} & \textbf{0.915} & \textbf{0.918} & \textbf{0.911} \\ 
\hline
\multirow{7}{*}{20\%}  & Transformer                                  & 0.631                  & 0.711                  & 0.622                  & 0.632                  &  & 0.623                  & 0.659                  & 0.614                  & 0.618                   \\
                       & Trans-mean                                   & 0.612                  & 0.742                  & 0.635                  & 0.641                  &  & 0.568                  & 0.594                  & 0.532                  & 0.553                   \\
                       & GRU-D                                        & 0.646                  & 0.733                  & 0.635                  & 0.648                  &  & 0.648                  & 0.698                  & 0.658                  & 0.672                   \\
                       & SeFT                                         & 0.357                  & 0.421                  & 0.381                  & 0.350                  &  & 0.342                  & 0.349                  & 0.346                  & 0.333                   \\
                       & mTAND                                        & 0.332                  & 0.369                  & 0.377                  & 0.373                  &  & 0.456                  & 0.492                  & 0.490                  & 0.490                   \\
                       & RAINDROP                                     & 0.665                  & 0.720                  & 0.679                  & 0.651                  &  & 0.713                  & 0.758                  & 0.725                  & 0.734                   \\
                       & Ours   & \textbf{0.699} & \textbf{0.850} & \textbf{0.705} & \textbf{0.692} & & \textbf{0.854} & \textbf{0.879} & \textbf{0.883} & \textbf{0.869}   \\ 
\hline
\multirow{7}{*}{30\%}  & Transformer                                  & 0.316                  & 0.264                  & 0.240                  & 0.190                  &  & 0.520                  & 0.552                  & 0.501                  & 0.484                   \\
                       & Trans-mean                                   & 0.425                  & 0.453                  & 0.370                  & 0.339                  &  & 0.651                  & 0.638                  & 0.679                  & 0.649                   \\
                       & GRU-D                                        & 0.451                  & 0.517                  & 0.421                  & 0.472                  &  & 0.580                  & 0.632                  & 0.582                  & 0.593                   \\
                       & SeFT                                         & 0.327                  & 0.279                  & 0.345                  & 0.280                  &  & 0.317                  & 0.310                  & 0.320                  & 0.280                   \\
                       & mTAND                                        & 0.275                  & 0.312                  & 0.306                  & 0.308                  &  & 0.347                  & 0.434                  & 0.363                  & 0.395                   \\
                       & RAINDROP                                     & 0.524                  & 0.609                  & 0.513                  & 0.484                  &  & 0.603                  & 0.681                  & 0.603                  & 0.619                   \\
                       & Ours   & \textbf{0.603} & \textbf{0.778} & \textbf{0.639} & \textbf{0.641} & & \textbf{0.732} & \textbf{0.817} & \textbf{0.735} & \textbf{0.754} \\ 
\hline
\multirow{7}{*}{40\%}  & Transformer                                  & 0.230                  & 0.074                  & 0.145                  & 0.069                  &  & 0.438                  & 0.446                  & 0.405                  & 0.402                   \\
                       & Trans-mean                                   & 0.257                  & 0.091                  & 0.185                  & 0.099                  &  & 0.487                  & 0.558                  & 0.542                  & 0.551                   \\
                       & GRU-D                                        & 0.464                  & 0.645                  & 0.426                  & 0.443                  &  & 0.477                  & 0.634                  & 0.445                  & 0.475                   \\
                       & SeFT                                         & 0.263                  & 0.299                  & 0.273                  & 0.223                  &  & 0.268                  & 0.241                  & 0.280                  & 0.233                   \\
                       & mTAND                                        & 0.194                  & 0.151                  & 0.202                  & 0.170                  &  & 0.237                  & 0.339                  & 0.264                  & 0.293                   \\
                       & RAINDROP                                     & 0.525                  & 0.534                  & 0.486                  & 0.447                  &  & 0.570                  & 0.654                  & 0.567                  & 0.589                   \\
                       & Ours  & \textbf{0.584} & \textbf{0.743} & \textbf{0.591} & \textbf{0.584} & & \textbf{0.682} & \textbf{0.766} & \textbf{0.691} & \textbf{0.695}\\ 
\hline
\multirow{7}{*}{50\%}  & Transformer                                  & 0.214                  & 0.027                  & 0.125                  & 0.044                  &  & 0.432                  & 0.520                  & 0.369                  & 0.419                   \\
                       & Trans-mean                                   & 0.213                  & 0.028                  & 0.125                  & 0.046                  &  & 0.464                  & 0.591                  & 0.431                  & 0.465                   \\
                       & GRU-D                                        & 0.373                  & 0.296                  & 0.328                  & 0.266                  &  & 0.497                  & 0.524                  & 0.425                  & 0.475                   \\
                       & SeFT                                         & 0.247                  & 0.159                  & 0.253                  & 0.182                  &  & 0.264                  & 0.230                  & 0.275                  & 0.235                   \\
                       & mTAND                                        & 0.169                  & 0.126                  & 0.170                  & 0.139                  &  & 0.209                  & 0.351                  & 0.230                  & 0.277                   \\
                       & RAINDROP                                     & 0.466                  & 0.445                  & 0.424                  & 0.380                  &  & 0.472                  & 0.594                  & 0.448                  & 0.476                   \\
                           & Ours  & \textbf{0.601} & \textbf{0.707} & \textbf{0.611} & \textbf{0.597} & & \textbf{0.667} & \textbf{0.750} & \textbf{0.686} & \textbf{0.674}\\
\hline
\end{tabular}
}
}
\vspace{-17pt}
\end{table}

% \begin{table}[ht]
% \label{noisy_experiment}
% \centering
% \caption{Classification performance of adding Gaussian noise with different variances on the UCI-HAR. }
% \begin{tabular}{l|l|c|c|c|c}
% \hline \hline
%                           & noisy std & Accuracy & Precision & Recall & F1 score \\ 
% \multirow{5}{*}{UCI-HAR} & 0         & 95.05    & 95.40     & 95.00  & 95.14    \\ 
%                           & 15        & 96.61    & 96.72     & 96.61  & 96.65    \\
%                           & 25        & 91.62    & 92.69     & 91.83  & 91.89    \\
%                           & 35        & 93.04    & 93.95     & 93.20  & 93.26    \\ 
%                           & 45        & 96.00    & 96.28     & 96.12  & 96.13    \\ \hline
% \multirow{5}{*}{WISDM}    & 0         & 99.33    & 98.97     & 99.24  & 99.10    \\ 
%                           & 15        & 99.36    & 98.92     & 99.00  & 98.96    \\ 
%                           & 25        & 99.17    & 98.61     & 98.86  & 98.73    \\ 
%                           & 35        & 99.14    & 98.53     & 98.73  & 98.63    \\ 
%                           & 45        & 99.33    & 99.04     & 99.23  & 99.13   \\
% \hline \hline
% \end{tabular}
% \end{table}

\begin{figure*}[htbp]
    \centering
 \includegraphics[width=7in]{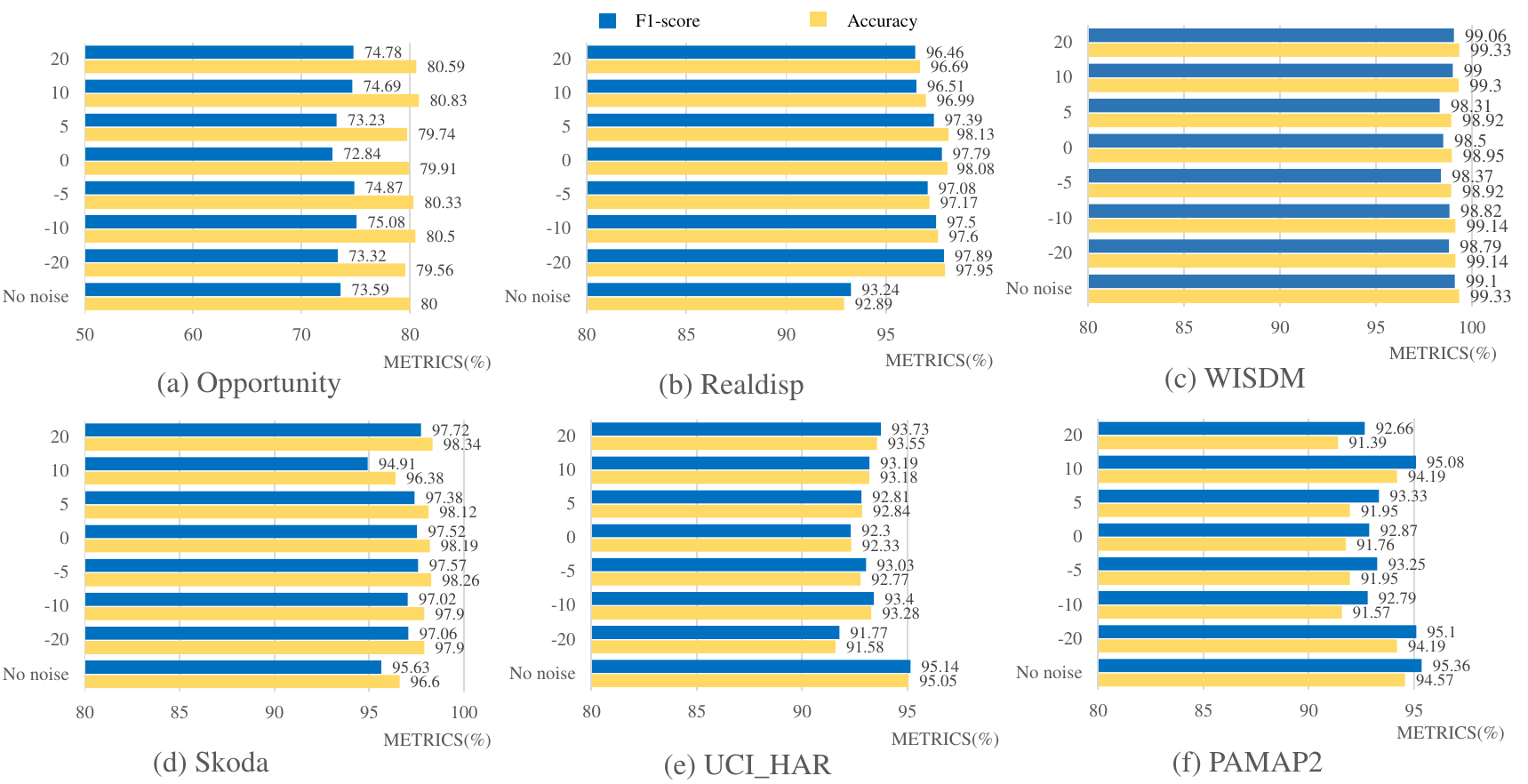}
    \caption{Classification performance on samples with different SNR.}
    \label{noisy_fig}
\end{figure*}

\subsection{Robust To Sensor Noise}
Sensor noise is inevitable in real-world scenarios and it can interfere the learning ability of artificial intelligence models. We evaluate the robustness of our model to sensor noise on six datasets by adding noise with different Signal-to-Noise Ratios (SNR) to the original signals. SNR is defined as:
\begin{equation}
\label{snr}
SNR = 10*\log_{10}\frac{P_{signal}}{P_{noise}}
\end{equation}
where \(P_{signal}\) is the power of signal, \(P_{noise}\) is the power of noise. 

We experiment with seven levels of noise intensity, corresponding to SNR values of -20, -10, -5, 0, 5, 10, and 20. Lower SNR corresponds to the higher noise intensity. When \(SNR\) is equal to zero, the power of the signal equals the power of noise. The results are shown in Fig. ~\ref{noisy_fig}. It is notable that:
%In this part, we experiment with seven groups of different intensity noise, SNR set up to -20, -10, -5, 0, 5, 10, 20 respectively. The smaller the SNR, the more noise intensity is added. When \(SNR=0\), the power of the signal is equal to the power of noise. The experiment results are shown in Fig. ~\ref{noisy_fig}. On Realdisp, realworld datasets, the model performance significantly improves after adding noise. Our model achieves the best performance, improved accuracy by 5.06\% and F1 score by 4.65\%  on the dataset adding noise with -20 SNR compared to the model trained with raw data. Although the accuracy and F1 score decrease slowly as the noise intensity decreases, they are still higher than the baseline. On the Realworld dataset, our model achieves the best performance when snr is equals to 10 with a 7.18\% increase in accuracy and a 10.65\% increase in F1-score. On Opportunity and Skoda  datasets, the performance of the model remained unchanged after adding noise of different intensities compared to the baseline The noises do not have a negative impact on the model. On the UCI-HAR and PAMAP2 datasets, model performance decrease slightly after adding noise. 
\begin{itemize}[left=0pt]
%\item The lower the SNR, the higher the noise intensity. When SNR is 0, the signal and noise have equal power.

\item On Realdisp and Realworld datasets, the model performance improves significantly after adding noise, especially when SNR is low (-20 or -10). This implies that the model is capable of extracting valuable information from noisy signals and improving discriminative features for classification.

\item On Opportunity and Skoda datasets, the model performance remains stable after adding noise, regardless of the SNR value. This indicates that the model is insensitive to noise and can maintain a high accuracy and F1-score on these datasets.

\item On UCI-HAR and PAMAP2 datasets, the model performance decreases slightly after adding noise, especially when SNR is high (10 or 20). This implies that the model may be affected by some high-frequency noise components that interfere with the signal patterns.
\end{itemize}

The experiment results demonstrate the robustness of our model to sensor noise on most datasets, as it can either improve or maintain its performance after adding noise. There are two main reasons for this robustness:
\begin{itemize}[left=0pt]
\item Signal is decomposed by wavelet transform into multiple sub-bands, each corresponding to a different frequency component. Noise is usually concentrated in the high-frequency sub-bands, while useful information is often found in the low-frequency sub-bands. By applying wavelet transform, our model can filter out most of the noise and concentrate on the informative aspects of the signal.
\item Parallel heterogeneous feature extractor is designed to extract features from each sub-signal at different scales and resolutions. It proves beneficial in enhancing the diversity and complexity of extracted signal patterns, enabling the model to learn more representative features for classification.
\end{itemize}

In conclusion, our model can effectively reduce the impact of noise on performance. Even in the presence of noise, useful features can be extracted from the data and  and achieve superior results.

% Table ~\ref{noisy_experiment} displays the classification performance of adding Gaussian noise with different variances on the UCI-HAR dataset. The accuracy and F1 score were used as the evaluation metrics. The results show that when there is no noise added, the accuracy and F1 score are 95.05\% and 95.14\%, respectively. As the intensity of noise increases, the accuracy and F1 score show varying degrees of decline, except for the case of 15 intensity, where the performance improves compared to the no noise case. The highest accuracy and F1 score are achieved when the intensity of noise is set to 15, with 96.61\% and 96.65\%, respectively. However, when the intensity of noise is set to 25, the performance decreases significantly with an accuracy of 91.62\% and an F1 score of 91.89\%. The classification performance slightly recovers when the intensity of noise increases to 35 with an accuracy of 93.04\% and an F1 score of 93.26\%. Finally, the classification performance achieves the second-best performance when the intensity of noise is set to 45, with an accuracy of 96.00\% and an F1 score of 96.13\%. Overall, the results demonstrate that adding Gaussian noise with an appropriate intensity can improve the classification performance, but excessive noise can have a negative impact on the classification accuracy and F1 score.

\begin{figure*}[!t]
    \centering
  \includegraphics[width=7in]{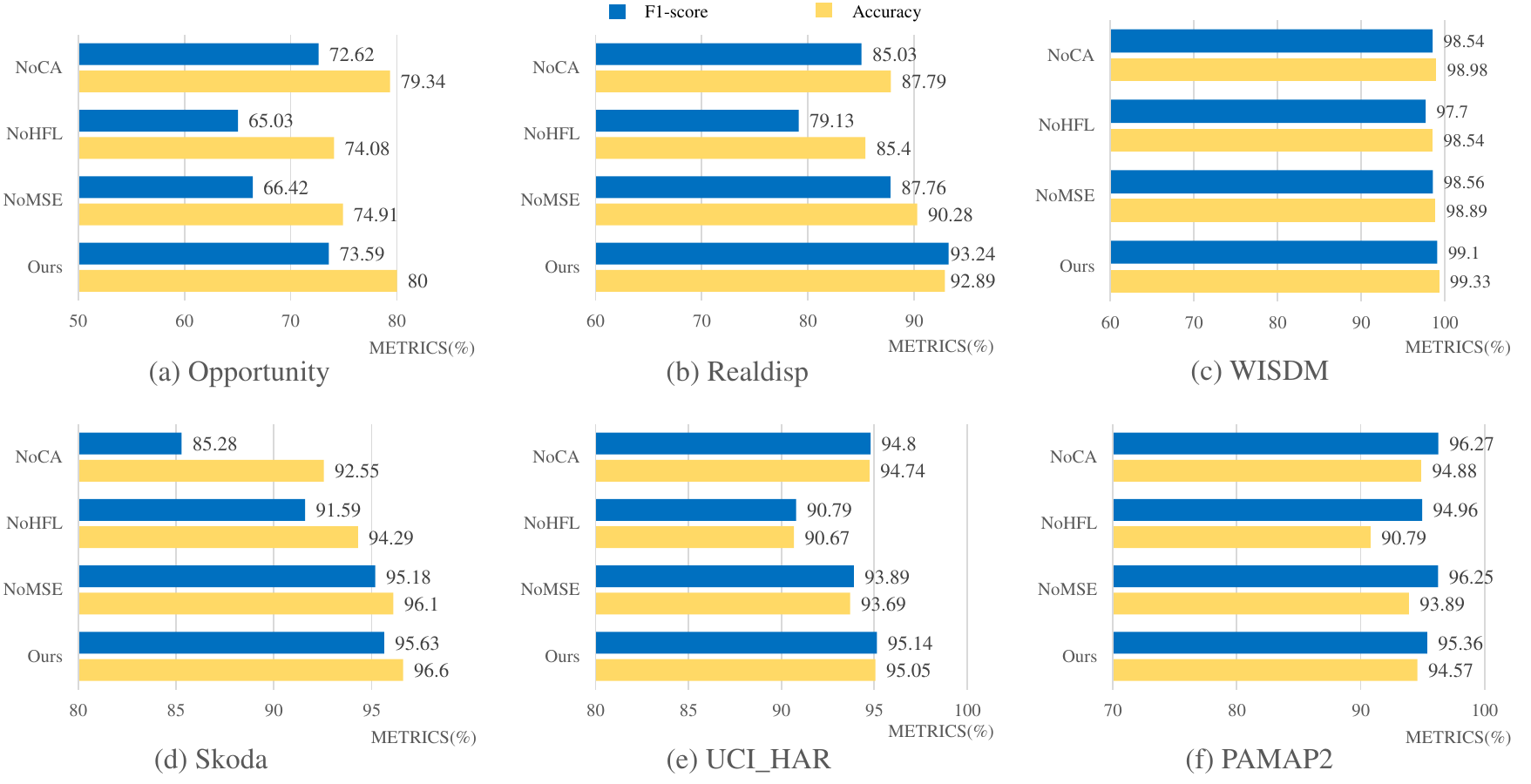}
    \caption{Ablation study for evaluating the importance of each module in \model~model.}
    \label{ablation_fig}
\end{figure*}

\subsection{Ablation Analysis}
In this subsection, we  assess the significance of each module of our model \model~by comparing it with simplified versions. The simplified versions are:

{\bf{NoMSE}}: This version removes the multilevel signals extractor module from \model. The original signal is used as the input for each part of the heterogeneous feature learner.

{\bf{NoHFL}}: This version replaces each part of the heterogeneous feature learner with a three-layer CNN structure. The signals extracted from the multilevel signals extractor with different resolutions are fed into the same feature learner.

{\bf{NoCA}}: This version removes the cross aggregation module from \model. The outputs of the heterogeneous feature learner are concatenated directly and then fed into the classifier.

Fig~\ref{ablation_fig} shows the results of ablation analysis. We can observe that:

\begin{itemize}[left=0pt]
\item On Opportunity, Realdisp, Realworld, Skoda and UCI-HAR datasets, \model~outperforms all simplified versions, which demonstrates the effectiveness of each module. Especially on Opportunity, Realdisp and Realworld datasets, the difference is more significant.
\item The performance of NoHFL is worse than \model~and other simplified versions on six datasets, which indicates that using heterogeneous feature learners for signals with different resolutions is beneficial for capturing multi-scale patterns.
\item Compared with the \model, the performance of NoMSE drops noticeably. This result confirms the usefulness of multilevel discrete wavelet decomposition for extracting multi-resolution features from the original signal.
\item The performance of NoCA are better than \model. This result shows that the cross aggregation module can enhance the feature interaction and representation by blending the features of different layers together.
\end{itemize}

In summary, each module of \model~plays a vital role in improving the model performance, especially the heterogeneous feature learner and the multilevel signals extractor.

\subsection{ Sensitivity Analysis}
\begin{figure*}[htbp]
    \centering
  \includegraphics[width=7in]{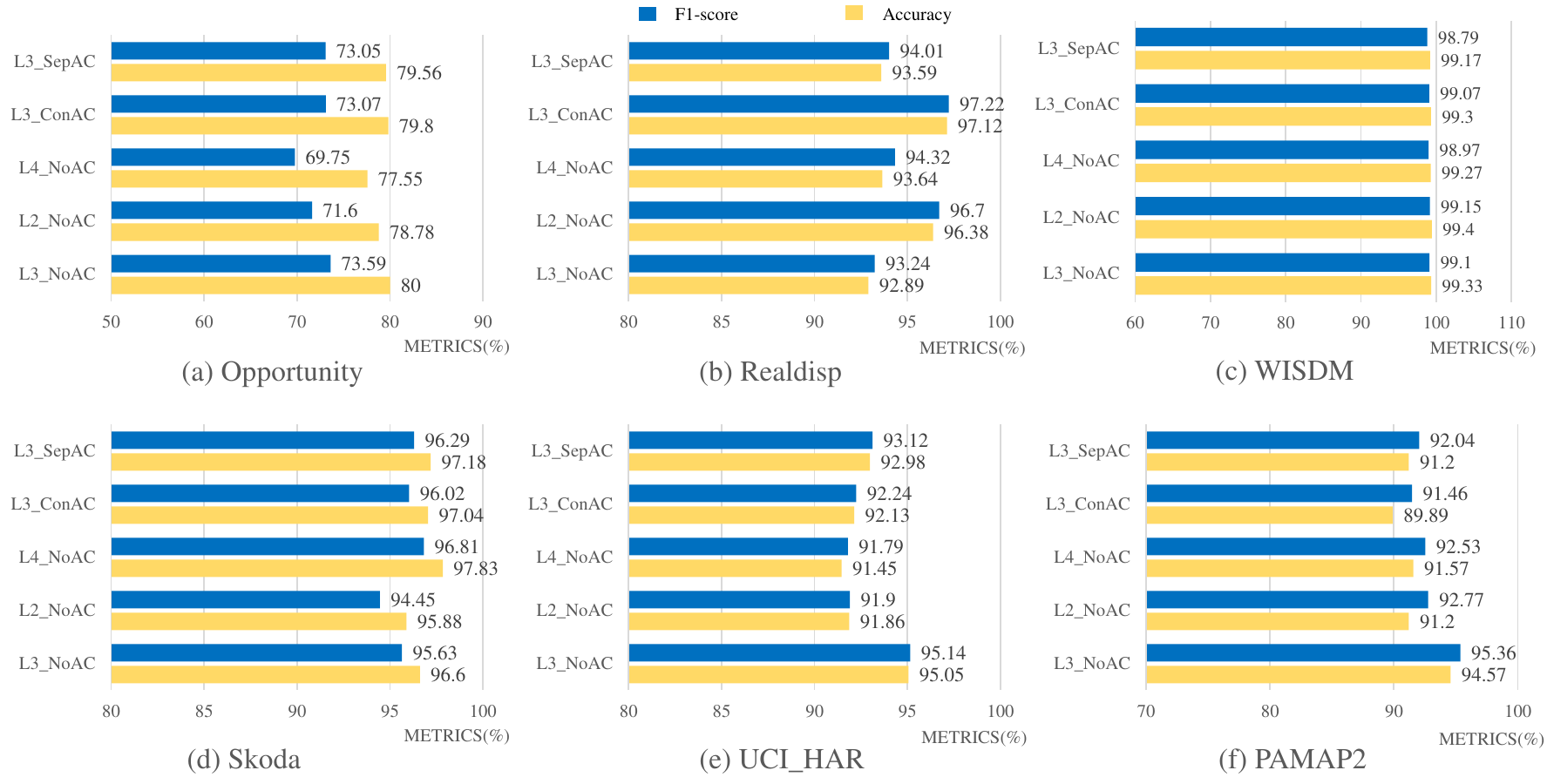}
    \caption{Sensitivity study for evaluating the importance of each module in \model~model.}
    \label{sensitivity_fig}
\end{figure*}

%In this subsection, we investigate how the performance of \model~is influenced by the levels of multilevel discrete wavelet decomposition and signals selection after  multilevel signals extractor. \(Lx\) represents the number of layers for multi-layer discrete wavelet decomposition is set to \(x\). \(NoAC\) represents the approximation coefficients do not input to extract feature.\(ConAC\) represents approximation coefficient and detail coefficient of the last level are concatenated as input of MLP in heterogeneous feature learner. \(SepAC\) represents the approximation coefficient and detail coefficient of the last level are input to a MLP module in heterogeneous feature learner, respectively.

%The results of sensitivity analysis are shown in Fig. \ref{sensitivity_fig}. On the Opportunity, UCI-HAR and PAMAP2 datasets, L3\_NoCA achieved the best results. The performance of L4\_NoCA is far superior to other models on the RealWorld dataset. And it outperforms other models on Skoda dataset slightly. L3\_ConAC get good performance on Realdisp, Opportunity and Realworld datasets. L3\_SepAC achieves good performance on Skoda and UCI-HAR dataset.

%In short, L3\_NoCA outperforms the other models on more datasets. Compared with L4\_NoCA, L3\_ConCA and L3\_SepCA, it has a smaller number of parameters. Hence, we use L3\_NoCA in the experiments.

In this subsection, we explore the sensitivity of \model~to the levels of multilevel discrete wavelet decomposition and components selection of multilevel signals extractor. Let \(Lx\) denotes the number of layers for multilevel discrete wavelet decomposition is set to \(x\). And the \(NoAC\) denotes that the approximation coefficient of last layer is not used as input for feature extraction. The \(ConAC\) represents that the approximation coefficient and detail coefficient of the last level are concatenated as input of MLP in heterogeneous feature learner. We use \(SepAC\) to denote that the approximation coefficient and detail coefficient of the last level are fed into separate MLP modules in heterogeneous feature learner.

The sensitivity analysis results are shown in Fig. \ref{sensitivity_fig}. We can observe that:
\begin{itemize}[left=0pt]
\item On Opportunity, UCI-HAR and PAMAP2 datasets, \(L3\_NoAC\) achieves the best results, which indicates that using three layers of multilevel discrete wavelet decomposition can capture the optimal features for these datasets. The approximation coefficient at last layer only contains limited information which is not useful for feature extraction.
\item On Realworld dataset, \(L4\_NoAC\) outperforms other versions by a large margin, which suggests that using four layers of multilevel discrete wavelet decomposition and ignoring the approximation coefficients can extract more informative features for this dataset.
\item On Skoda dataset, \(L3\_SepAC\) and \(L4\_NoAC\) perform slightly better than other models, which implies that using three layers of multilevel discrete wavelet decomposition and separating the approximation coefficient and detail coefficient of the last level can enhance the feature representation for this dataset.
\item On Realdisp dataset, \(L3\_ConAC\) performs better than other models except \(L3\_NoAC\), which shows that using three layers of multilevel discrete wavelet decomposition and concatenating the approximation coefficient and detail coefficient of the last level can improve the feature interaction for this dataset.
\end{itemize}

In summary, \(L3\_NoAC\) performs well on most datasets and has fewer parameters than other models. Therefore, we use \(L3\_NoAC\) as our default setting in the experiments.

\subsection{Discussion}

In this paper, we tackle the research question of designing a novel architecture for sensor-based HAR that can overcome noise and missing values in sensor data. Our hypothesis is that using multilevel discrete wavelet decomposition, heterogeneous feature learners, and cross aggregation techniques can improve the feature extraction and representation from multi-resolution signals, and thus enhance the classification performance.

To test our hypothesis, we propose a novel structure called \model. It consists of three main modules: multilevel signals extractor, heterogeneous feature learner, and cross aggregation module. The multilevel signals extractor appl\textbf{}ies multilevel discrete wavelet decomposition to the original signal, producing multiple sub-signals with different resolutions and frequency components. This technique can effectively isolate noise from useful information. %, as noise is usually concentrated in the high-frequency sub-signals, while useful information is concentrated in the low-frequency and intermediate-frequency sub-signals. Moreover, this technique can also deal with missing values in sensor data, as it can reconstruct the signal from the available sub-signals.
The heterogeneous feature learner consists of multiple subnets with different structures, each designed to extract features from a specific sub-signal. This module can capture the diversity and complexity of the signal patterns at different scales and resolutions, and learn more representative features for classification. The cross aggregation module is responsible for blending multi-resolution features together, enhancing the feature interaction and representation.

Our model undergoes evaluation on seven public datasets, covering diverse sensor types, activities, and environments. And 28 remarkable methods are chosen as baseline, such as DeepConvLSTM, Att. Model, and CraphConvLSTM, etc. Our model outperforms these methods on most datasets, demonstrating its effectiveness and robustness. Additionally, we conduct ablation and sensitivity analyses to access the importance of each module in our model and the influence of different parameters.

Nevertheless, our model exhibits certain limitations that warrant attention in future work. First, the convergence behavior of the subnets in the heterogeneous feature learner may vary depending on the complexity and resolution of the signals. Therefore, we plan to explore methods to monitor and adjust the training status of each subnet, such as visualization or independent verification techniques. Second, the challenge of sample imbalance is still not well solved by our model. In real-world scenarios, some activities may be more frequent or dominant than others, resulting in skewed data distribution. This may affect the model’s ability to accurately recognize rare or minor activities. Therefore, we intend to optimize our model by improving the model structure and the loss function, as well as applying data augmentation or resampling techniques.

\section{CONCLUSION}
\label{conclusion}
%The growing popularity of compact intelligent electronic devices underscores the profound practical significance of sensor-based behavior recognition. However, this field presents formidable challenges due to the pervasive presence of noise and missing values within the data. In response to these challenges, we introduce a novel algorithm, the Multilevel Heterogeneous Neural Network (\model). Our approach leverages multistage discrete wavelet decomposition to mitigate the impact of noise and interpolation. Furthermore, we enhance the feature set through the deployment of heterogeneous feature learners and cross fusion techniques. Our experimental results unequivocally demonstrate the superiority of the proposed model over existing state-of-the-art methods, showcasing its robustness to noise and missing values.

The growing popularity of compact intelligent electronic devices underscores the profound practical significance of sensor-based behavior recognition. However, this field presents formidable challenges due to the pervasive presence of noise and missing values within the data. In response to these challenges, we introduce a novel algorithm, the Multilevel Heterogeneous Neural Network (\model). Our approach leverages multilevel discrete wavelet decomposition to mitigate the impact of noise and interpolation. Furthermore, we enhance the feature set through the deployment of heterogeneous feature learners and cross fusion techniques. We evaluate our model on seven public datasets and compare it with 28 state-of-the-art methods, covering various types of sensors, activities, and environments. The experiment results unequivocally demonstrate the efficacy of our model over existing methods, showcasing its robustness to noise and missing values. We have also performed ablation and sensitivity analyses to evaluate the contribution and impact of each module in our model. Our future work will focus on addressing some limitations of our model, such as the convergence behavior of the subnets in the heterogeneous feature learner and the challenge of sample imbalance in real-world scenarios. We hope that our work serves as inspiration for further research in the development of sensor-based behavior recognition.

% \section*{Acknowledgments}
% This should be a simple paragraph before the References to thank those individuals and institutions who have supported your work on this article. 

\bibliographystyle{IEEEtran}
\bibliography{IEEEabrv, reference}

\end{document}